\documentclass[twocolumn]{aastex62}

\usepackage{diagbox}

\newcommand{\Teff}{$T_{\rm{eff}}$}

\newcommand{\Msun}{$M_{\odot}$}

\newcommand{\Lstar}{$L_{\star}$}

\newcommand{\accunit}{$_{\odot}$ yr$^{-1}$}

\newcommand{\rev}{ }
\newcommand{\newrev}{ }
\newcommand{\newnewrev}{ }

\newcommand{\sect}{Sect.\,}

\def\arcsec{\hbox{$^{\hbox{\rlap{\hbox{\lower4pt\hbox{$\,\prime\prime$}}}\hbox{$\frown$}}}$}}
          
\graphicspath{{./}{figures/}}

\shorttitle{}
\shortauthors{}

\begin{document}

\title{An Improved HR Diagram for the Orion Trapezium Cluster}

\author{Min Fang}
\affiliation{Department of Astronomy, University of Arizona, 933 North Cherry Avenue, Tucson, AZ 85721, USA}
\affiliation{Earths in Other Solar Systems Team, NASA Nexus for Exoplanet System Science}
\author{Jinyoung Serena Kim}
\affiliation{Department of Astronomy, University of Arizona, 933 North Cherry Avenue, Tucson, AZ 85721, USA}
\affiliation{Earths in Other Solar Systems Team, NASA Nexus for Exoplanet System Science}
\author{Ilaria Pascucci}
\affiliation{Department of Planetary Sciences, University of Arizona, 1629 East University Boulevard, Tucson, AZ 85721, USA}
\affiliation{Earths in Other Solar Systems Team, NASA Nexus for Exoplanet System Science}
\author{D\'aniel Apai}
\affiliation{Department of Astronomy, University of Arizona, 933 North Cherry Avenue, Tucson, AZ 85721, USA}
\affiliation{Department of Planetary Sciences, University of Arizona, 1629 East University Boulevard, Tucson, AZ 85721, USA}
\affiliation{Earths in Other Solar Systems Team, NASA Nexus for Exoplanet System Science}

\begin{abstract}

  In this paper, we present a study of the Trapezium cluster in Orion. We analyze flux-calibrated VLT/MUSE spectra of 361 stars to simultaneously measure  the spectral types, reddening, and the optical veiling due to accretion. {\newnewrev We find that the extinction law from \citet{1989ApJ...345..245C} with a total-to-selective extinction value of $R_{\rm V}=$5.5 is more suitable for this cluster.}  For 68\% of the sample the new spectral types are consistent with literature spectral types within 2 subclasses, but as expected, we derive systematically later types than the literature by one to two subclasses for the sources with significant accretion levels.  Here we present an improved   Hertzsprung-Russell (H-R) diagram of the Trapezium cluster, in which the contamination by optical veiling on spectral types and stellar luminosities has been properly removed.  A comparison of the locations of the stars in the H-R diagram with the non-magnetic and magnetic pre-main sequence evolutionary tracks indicates an age of 1--2~Myr. The magnetic pre-main sequence evolutionary tracks can better explain the luminosities of the low-mass stars. In the H-R diagram, the cluster exhibits a large luminosity spread ($\sigma$(Log~$L_{\star}/L_{\odot})\sim$0.3).  By collecting a sample of 14 clusters/groups with different ages, we find that the luminosity spread tends to be constant  ($\sigma$(Log~$L_{\star}/L_{\odot})\sim$0.2--0.25) after 2~Myr, which suggests that age spread is not the main cause of the spread. {\newnewrev There are $\sim$0.1~dex larger luminosity spreads for the younger clusters, e.g., the Trapezium cluster, than the older clusters, which can  be explained by the starspots, accretion history and circumstellar disk orientations.}

\end{abstract}




\section{Introduction} \label{sec:introduction}

T~Tauri stars (TTSs) are low-mass pre-main sequence (PMS) stars, characterized by large variability and strong emission lines \citep{1945ApJ...102..168J}. They are classified as either classical T Tauri stars (CTTSs) or weak-lined T Tauri stars (WTTSs) based on the strength of their H$\alpha$ emission line (e.g., \citealt{1988cels.book.....H}): CTTSs exhibit broad and strong  H$\alpha$ emission lines attributed to accretion, while WTTSs show narrow and weak H$\alpha$ emission lines due to chromospheric activity \citep{2003ApJ...582.1109W}. 

For CTTSs, accretion produces not only strong permitted emission lines, such as the hydrogen Balmer series, but also ultraviolet/optical excess continuum emissions \citep{1998ApJ...509..802C}, which veil the  photospheric absorption features in  stellar spectra \citep{2004ApJ...616..998W}. Without accounting for this veiling effect, the {\newnewrev estimated} spectral types (of the veiled spectra) tend to appear earlier than the actual spectral types (e.g. \citealt{2014ApJ...786...97H}). This effect further complicates the analysis of  the extinction and luminosity of young stars. Brightness variability of young stars is an another complicating factor  when deriving extinction and stellar luminosity from broad-band photometry observed at different times, and this variability affects both CTTSs and WTTSs (see extreme cases like the CTTS GI~Tau in \citealt{2018ApJ...852...56G} and the WTTS LkCa~4 in \citealt{2017ApJ...836..200G}). 

Both spectral type (SpT), which is converted to an effective temperature ($T_{\rm eff}$) via a SpT-$T_{\rm eff}$ relation (e.g., \citealt{2013ApJS..208....9P}), and stellar luminosity ($L_{\star}$) are necessary
to place young stars in the H-R diagram. From their locations in the H-R diagram compared to  theoretical isochrones, ages and  masses of young stars are  determined  (e.g. \citealt{1997AJ....113.1733H}). Errors in measuring SpT and $L_{\star}$ could result in luminosity scatter in H-R diagram, which may be wrongly interpreted as an age spread \citep{2008ASPC..384..200H,2014prpl.conf..219S}. Simultaneous measurements of extinction, ultraviolet/optical excess continuum emission, SpT, and $L_{\star}$ have been performed on flux-calibrated optical spectra of  young stars in nearby low-mass star-forming regions, e.g. Taurus, Lupus, Chamaeleon I, etc. \citep{2014ApJ...786...97H,2014A&A...561A...2A,2017A&A...600A..20A,2016A&A...585A.136M,2017A&A...604A.127M}. In this work, we perform a similar study for the massive Orion Trapezium Cluster. 

The  Trapezium Cluster in the heart of the Orion nebula, at an age of $\sim$1--2\,Myr \citep{1997AJ....113.1733H,2010ApJ...722.1092D}, is one of the nearest \citep[388$\pm$10~pc][]{2017ApJ...834..142K} massive clusters, and has been an excellent laboratory for studying critical aspects of star and planet formation, e.g., initial mass function, cluster formation, protoplanetary disk evolution \citep{1997AJ....113.1733H,2000ApJ...540..236H,2000ApJ...540.1016L,2004ApJ...610.1045S,2010ApJ...722.1092D,2012ApJ...748...14D,1998AJ....116.1816H,2000AJ....120.3162L,2018ApJ...860...77E}.

Here, we simultaneously derive the extinction, veiling, SpT, and $L_{\star}$ of young stars in the Trapezium Cluster,  based on {\newnewrev wide field spectroscopic data generated by the integral field spectrograph MUSE on the VLT} \citep{2014Msngr.157...13B}. In \S\ref{Sect:data}, we describe our data and spectral extraction. In \S\ref{Sect:typing}, we present our analysis, which includes spectral typing and measurements of stellar properties. {\rev In \S\ref{Sect:extlaw}, we discuss the suitable extinction law in the Trapezium cluster.}  In $\S$\ref{SECT:Comparison}, we compare our measurements with those in the literature, and construct the H-R diagram of the cluster in \S\ref{Sect:HRD}.  We discuss the implications of our results in $\S$\ref{SECT:discussion} and summarize our findings in $\S$\ref{SECT:summary}.

\begin{figure*}
\begin{center}
\includegraphics[angle=0,width=1.8\columnwidth]{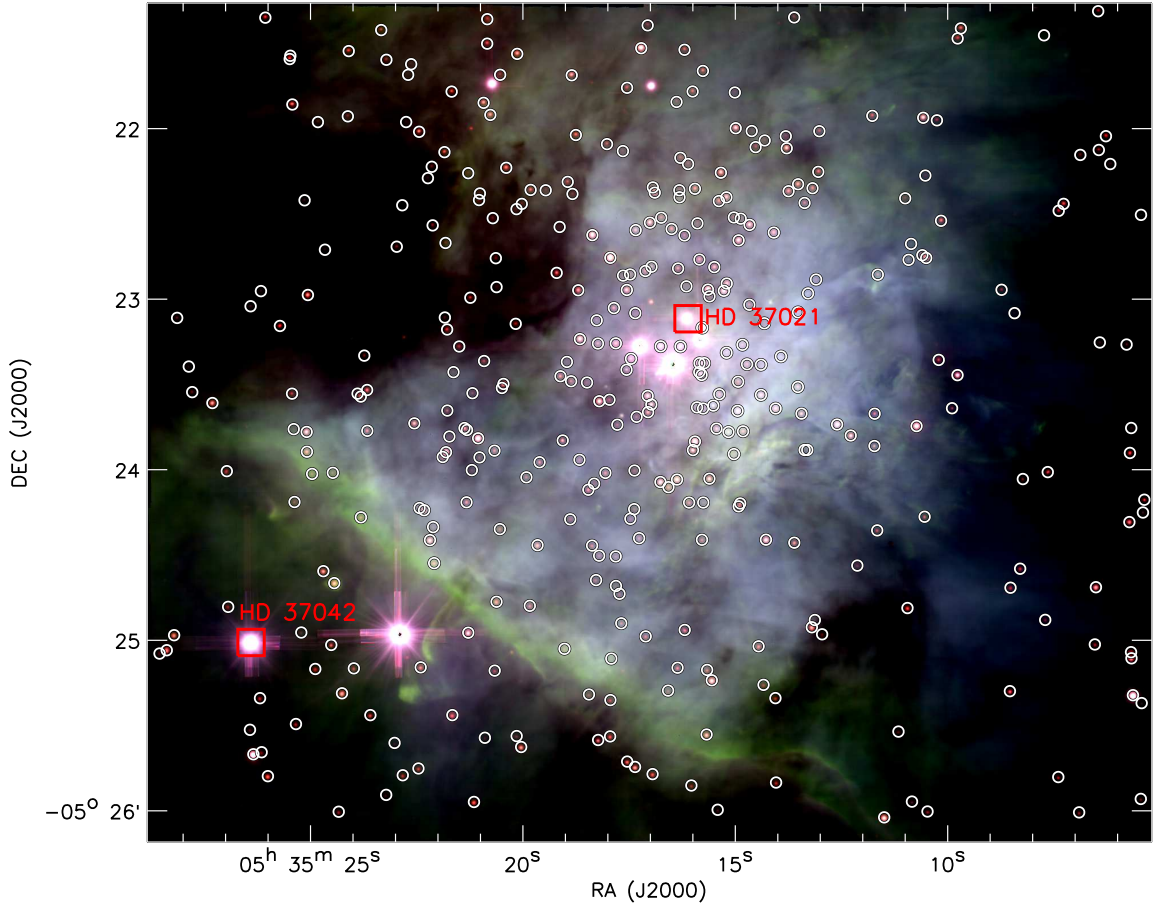}
\caption{Three-color image of the  Trapezium Cluster  created with the MUSE data integrated over 5,000-6,000\AA (blue),  6,000-7,000\AA(green), and  7,500-8,500\AA (red). The white open circles mark the sources for which  spectra are extracted for the study in this work. {\rev Red squares show the two sources, HD 37042 and HD 37021, which were used} to set and to verify the flux calibration levels.} \label{Fig:ONC}
\end{center}
\end{figure*}

\section{Data}\label{Sect:data}
{\newrev The data were taken with the  Multi-Unit
Spectroscopic Explorer (MUSE). MUSE is the second-generation Integral Field Unit (IFU) mounted on the Very Large Telescope (VLT) at the Nasmyth B focus of the Yepun (the VLT UT4 telescope) in Paranal, Chile. The observations were carried out during the first commissioning run \citep{2014Msngr.157...13B}, under photometric conditions with the DIMM (Differential Image Motion Monitor) seeing ranging from 0\farcs67 to 1\farcs25, and airmass changing from 1.067 to 1.483 \citep{2015A&A...582A.114W}.  The observations are performed in Wide Field Mode with a field of view (FoV) covering 1$^{\arcmin}\times$1$^{\arcmin}$ and a pixel size of 0\farcs2$\times$0\farcs2. A total of 30 pointings over a 6-by-5 mosaic are taken to cover the Orion Trapezium Cluster with a field of 5\farcm9$\times$4\farcm9.}

The data have been reduced with the public MUSE pipeline by \cite{2015A&A...582A.114W}, which provides two versions of wavelength and flux-calibrated IFU datacubes with a sampling of 1.25~\AA\ and 0.85~\AA\ and a spectral resolution of $\sim$3,000. These are publicly available\footnote{The reduced  data are downloadable via the link \url{http://MUSE-vlt.eu/science/m42/}}. In this work, we take advantage of the reduced dataset with a sampling of 1.25~\AA\ to characterize the properties of young stars in the cluster.

We extract the spectra by performing  aperture photometry on each wavelength plane of the reduced IFU datacubes. For isolated sources, we adopt an aperture of 0\farcs8 and a sky annulus from 2\farcs0 to 3\farcs0,  while an aperture of 0\farcs6 is used for objects that have nearby sources with separations less than 3\farcs0.  For objects that appear near bright sources, we adopt both a smaller aperture (0\farcs6) and smaller sky annulus (0\farcs8 to 1\farcs6), in order to reduce the contamination. We apply wavelength-dependent aperture correction for each source. Due to  variations of the  seeing and airmass during the observations, we have divided the field into 30 sub-fields, and obtained the aperture corrections in each sub-field. The aperture corrections for individual sources are obtained using the isolated bright point sources by comparing their fluxes extracted using the apertures used for the targets  and a large aperture of 2\farcs0 and sky annulus from 3\farcs0 to 4\farcs0. Figure~\ref{Fig:ONC} shows the distribution of sources with extracted spectra within the cluster, and Table~\ref{tab:tabe_ONC} list these sources.

The relative flux calibration of MUSE data is expected to be accurate to at least 5\% \citep{2015A&A...582A.114W}. We check the absolute flux calibration using the bright HD~37042, see  Figure~\ref{Fig:ONC}. A comparison between the MUSE spectra for HD~37042 and the best-fit model atmosphere suggests that the flux is underestimated only by $\sim$5\%; see the detailed description in Appendix~\ref{Appen:cali}. Therefore, we adjust the flux calibration levels  using the ratio between the observed spectra and the best-fit model of HD~37042.
We use the star HD~37021, {\rev adopting $T_{\rm eff}=18500$\,K}, in the field to test our flux calibration.  We correct the spectrum of HD~37021 using the wavelength-dependent factor obtained from   HD~37042, and compare it with its best-fit model atmosphere. The median difference between them is about 2\%. %

\section{Spectral fitting and stellar properties}\label{Sect:typing}

 \subsection{Pre-main sequence spectral templates}\label{Sect:template}
{\newrev In Appendix~\ref{Appen:template}, we list the WTTSs with X-shooter spectra that we used as PMS spectral templates. These sources are  from the $\eta$~Cha cluster, the TW Hydrae Association, the Lupus star-forming region, the $\sigma$~Ori cluster, Upper Sco, Taurus, and the Cha I star-forming region. Most of the sources we used for this study have been {\newnewrev already} published  \citep{2013A&A...551A.107M,2017A&A...605A..86M,2018A&A...609A..70R}}.

We extract the spectra of these sources from the X-shooter phase III data archive. The X-shooter phase III spectral data are flux calibrated,  but not corrected for the slit losses. We re-calibrate these spectra using the standard star observed closest in time to each source and with the similar airmass to the source and the same instrumental setting as the source. For each standard star, we obtain the BOSZ Kurucz model atmosphere \citep{2012AJ....144..120M} corresponding to its spectral type. Next, we fit its broad-band photometry  using the aforementioned model with two free parameters, extinction and stellar angular radius as in \citet{2009A&A...504..461F,2013ApJS..207....5F}. Then, we shift and rotationally broaden the best-fit model atmosphere and degrade it to the X-shooter spectral resolution of the standard star. In the optical bands (UVB and VIS arm), the ratio between the X-shooter phase III spectrum and model spectrum  is fitted with a 5-order polynomial function, from which we obtain the flux correction as a function of wavelengths. Implementing the correction, we  obtain the flux-corrected spectra of the source and the standard star in the same wavelength. We divide the flux-corrected spectrum of the standard star by its model spectrum to obtain the telluric spectrum. The telluric spectrum is then scaled to match the telluric absorption in the spectrum of the source, and the scaled telluric spectrum is then applied for  telluric correction of the spectrum of the source by dividing the flux-corrected spectrum by the scaled telluric spectrum. In the near-infrared bands (NIR arm), we directly implement the ratio between the  X-shooter phase III spectrum and model spectrum of the standard star to correct the X-Shooter phase III spectrum of the source. In this way, we can correct for both the telluric absorption and flux.

All of the WTTSs used as stellar templates  have been previously classified in the literature (see Appendix~\ref{Appen:template}). In the Appendix~\ref{Appen:template}, we also describe how we determine the spectral types of individual sources used in this work as well as their visual extinction. The majority of the sources  have  negligible extinction based on their photometry in optical and near-infrared bands. For the sources with measurements of extinction, we list their visual extinction and deredden their X-shooter spectra using the visual extinction with the extinction law from \citet{1989ApJ...345..245C}, adopting a total-to-selective extinction value of $R_{\rm V}=$3.1. All the X-shooter templates have been then degraded to the spectral resolution ($\sim$3000) of the MUSE spectra. For the X-shooter spectra with the same spectral types, we obtain one mean spectrum from their degraded spectra and use it as the X-shooter template with that spectral type.  {\newnewrev This set of the X-shooter templates have been employed to do the spectral classification of young stars in the North America and Pelican Nebulae Region \citep{2020arXiv200911995F}.}

\subsection{Spectral typing}
\begin{figure*}
\begin{center}
\includegraphics[angle=0,width=1\columnwidth]{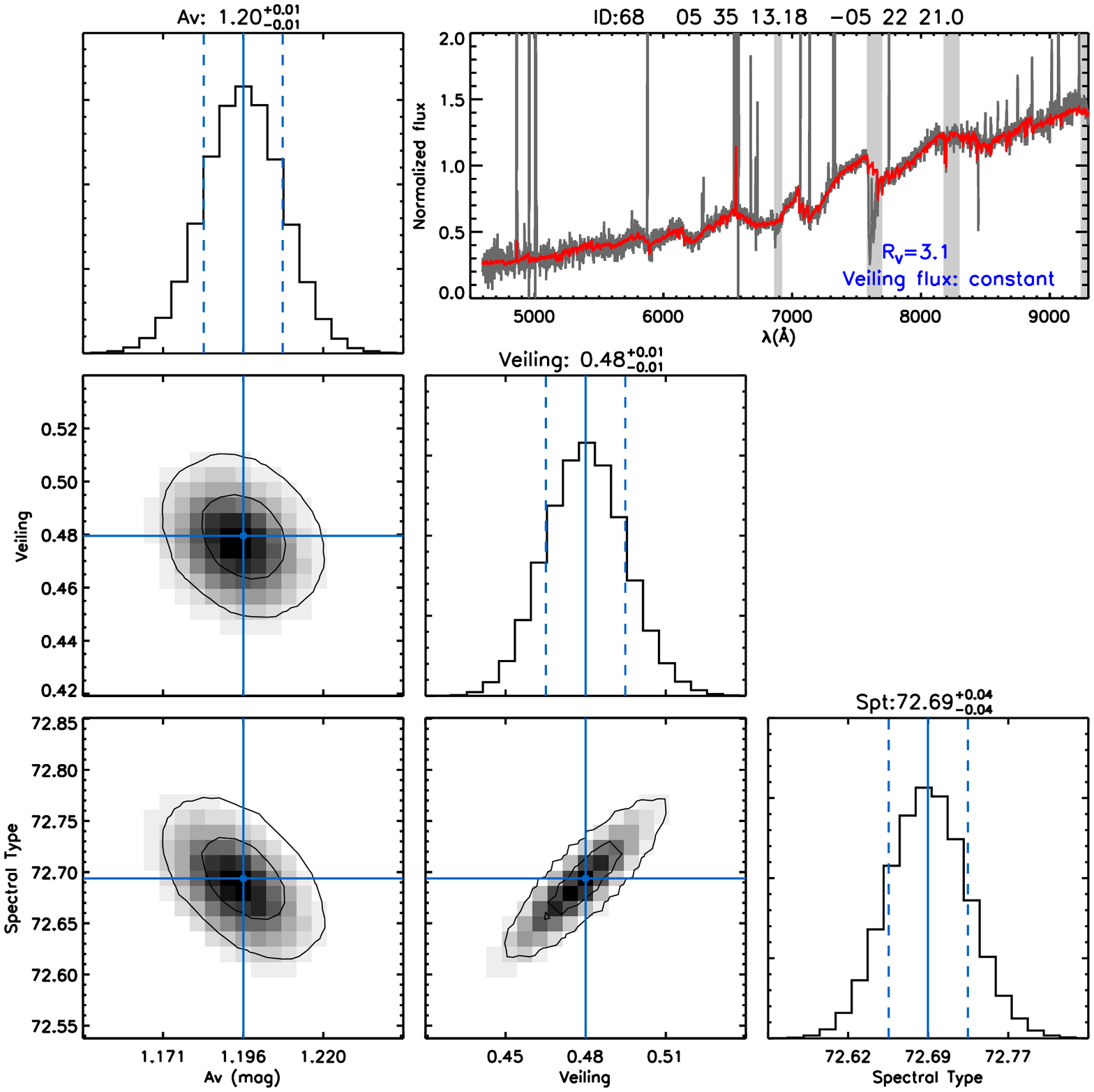}
\includegraphics[angle=0,width=1\columnwidth]{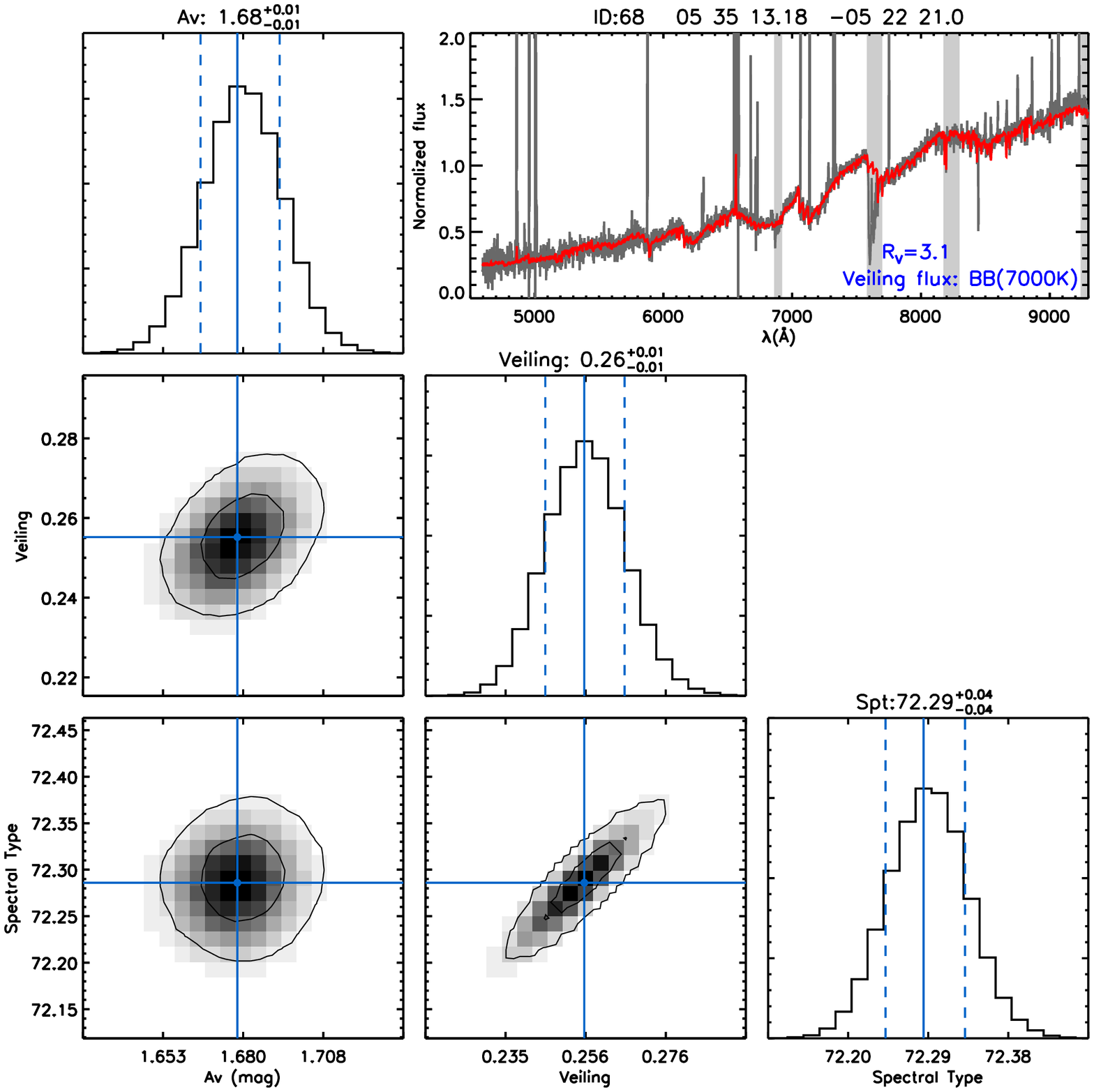}
\includegraphics[angle=0,width=1\columnwidth]{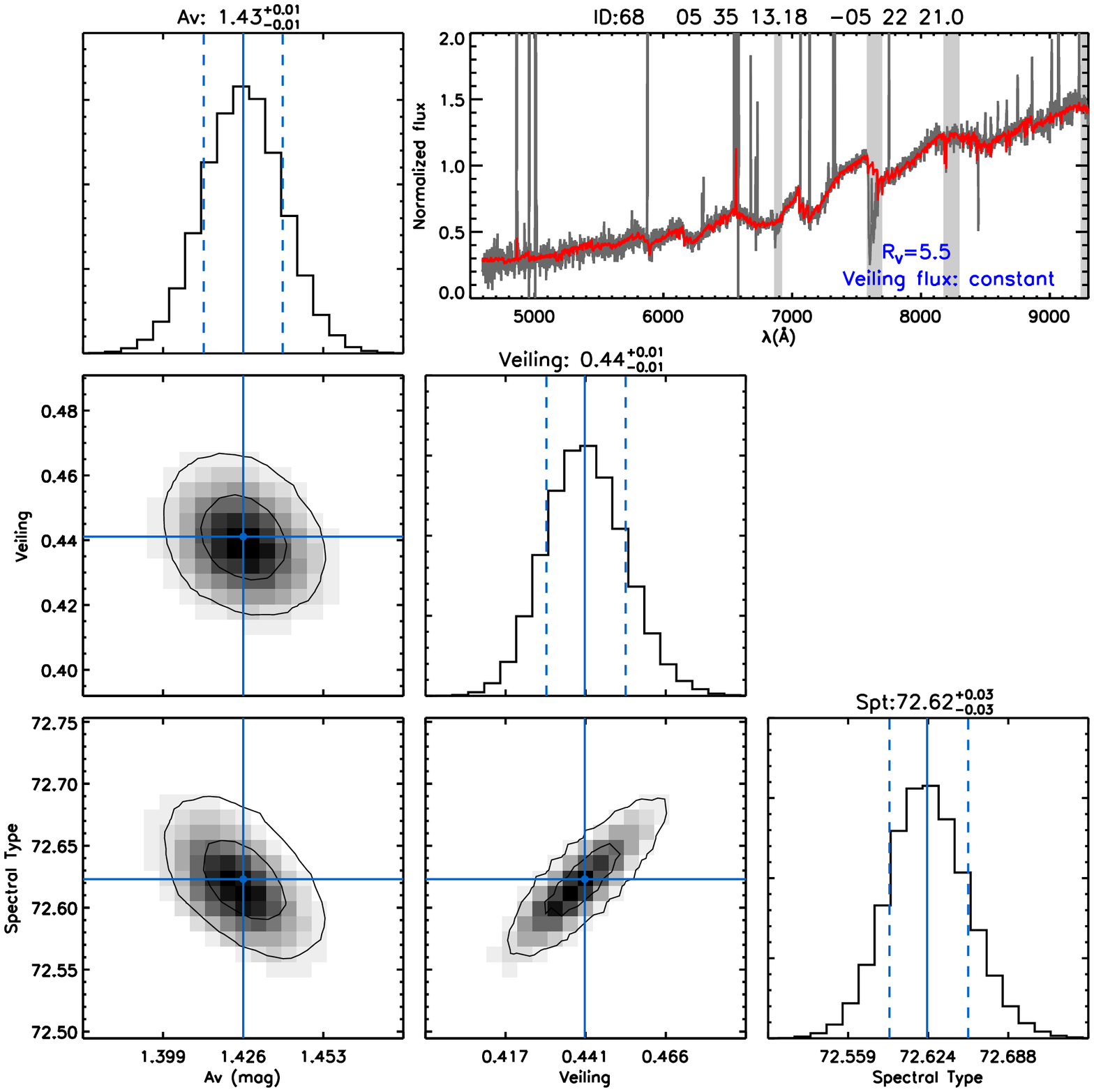}
\includegraphics[angle=0,width=1\columnwidth]{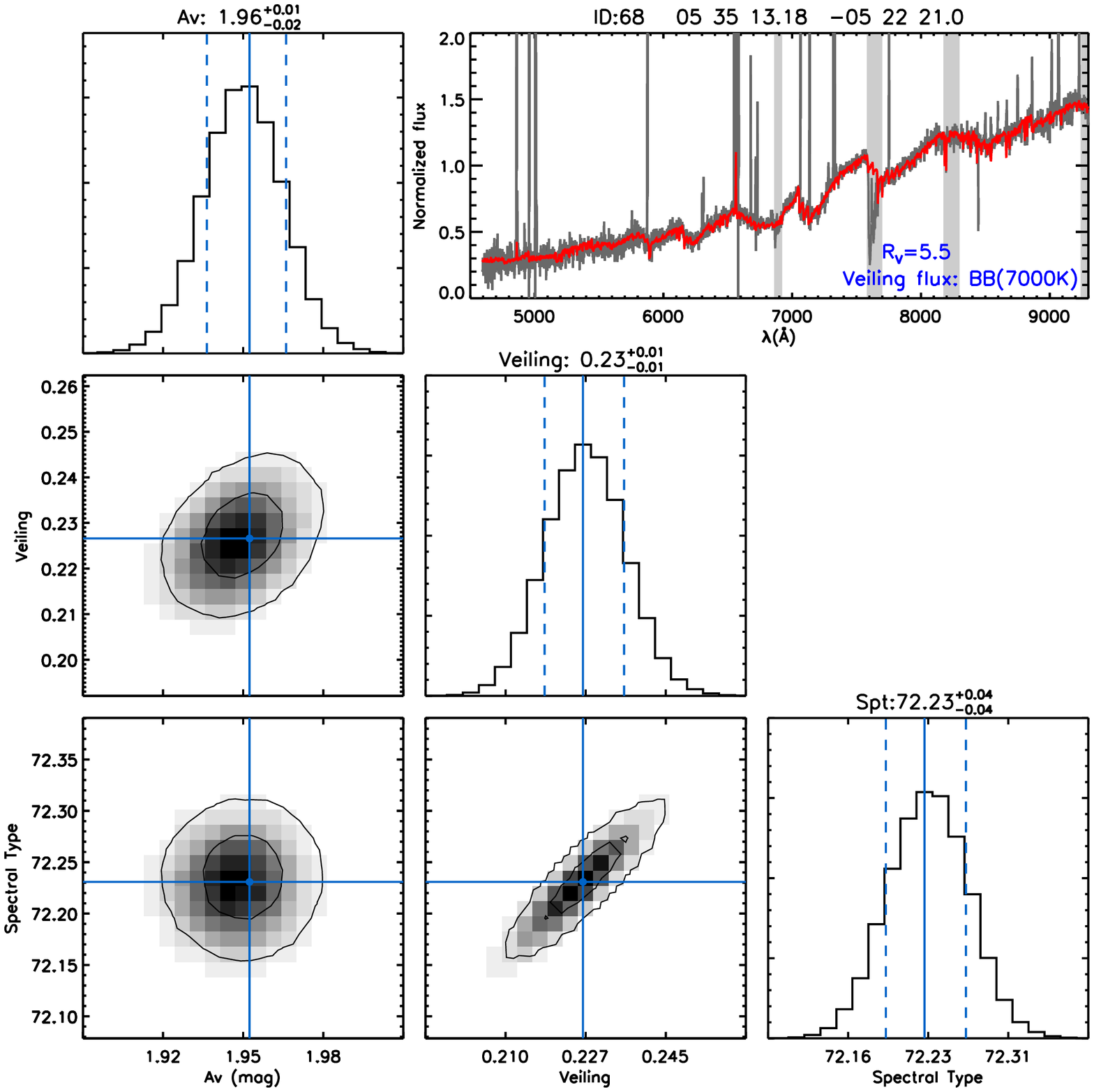}
\caption{Corner plot showing the posterior distributions from the MCMC fit of the MUSE spectra for Source 68 for the four sets of parameters with different $R_{\rm V}$ and the shapes of Accretion Continuum Spectrum. The vertical dashed lines are the 16 and 85 percentiles, respectively. The solid lines indicate the medians of the posterior distributions. The best-fit templates (red) are overlapped on the observed MUSE spectra (grey).} \label{Fig:Corner_plot}
\end{center}
\end{figure*}

We determine the spectral types of the targets with MUSE spectra by fitting their spectra over $\sim$5,500--9,000\,\AA\ with the spectral templates constructed with X-shooter spectra. {\rev For one source (Source 133) in our sample with an early-G type, earlier than the ones of X-shooter templates, we fit it with the model atmosphere with a solar abundance and surface gravity $log~g$=4.0 from  \cite{2013A&A...553A...6H}.} Our fitting procedure  employs three free parameters: the spectral type, the extinction ($A_{\rm V}$) in the $V$ band, and the veiling due to  accretion.  Given that the spectral resolution of MUSE is relatively low ($\sim$3,000), we do not consider the rotational broadening of the stars in the fitting. A Markov Chain Monte Carlo (MCMC) procedure is used to find the best-fit model parameters. When running the MCMC procedure, the used spectral templates are interpolated {\rev over spectral types} using the grid of our spectral templates. Then an excess flux is added to the spectral template, which is parameterized as $r_{7465}=\frac{F_{\rm excess,\, 7465}}{F_{\rm phot, \,7465}}$, where $r_{7465}$ is the veiling at 7465\,\AA, $F_{\rm excess, \, 7465}$  is the excess flux at \,7465\AA, and $F_{\rm phot,\, 7465}$ is the photospheric emission at  7465\,\AA.  We choose 7465\,\AA\ since at this wavelength there are no telluric absorption features and no strong absorption photospheric features.  We adopt two types of the Accretion Continuum Spectrum. One is the same as in \cite{2014ApJ...786...97H}, and is approximated by a constant over 5500-9000\,\AA, and the other is assumed to be a blackbody spectrum with $T=7000$~K as in \cite{2010ApJ...722.1092D}.  The veiled spectral template is then reddened with a visual extinction $A_{\rm V}$ using  the extinction law from \citet{1989ApJ...345..245C} with $R_{\rm V}=$3.1 (typical of interstellar medium dust) and 5.5. The resulting template is normalized at 7465\,\AA, and compared with the normalized spectrum of each object. In Fig.~\ref{Fig:Corner_plot}, we show the posterior distributions with the best-fit parameters and their uncertainties for Source 68 as an example.  With different assumptions, its spectral type, $A_{\rm V}$  and $r_{7465}$ range from M2.2 to M2.7, 1.20 to 1.96 mag, and 0.23 to 0.48, respectively.

In total, we measure spectral types for 361 sources. {\rev Among them, 360 are determined with X-Shooter templates, and one (Source 133) uses the model atmosphere.} There is no significant systematic difference (median $\Delta$SpT$\lesssim$0.15~subclass) in the spectral type derived based on  different extinction $R_{\rm V}$ and the shapes of Accretion Continuum Spectrum. For more than 96\% of the sources the spectral types agree well with each other within one subclass. Larger differences are seen mostly among the K or G type sources and/or with large veiling. For the constant Accretion Continuum Spectrum case,  there are 3 sources with differences around 1 subclasses between $R_{\rm V}=$3.1 and 5.5. Two of them are G spectral type sources, and one is mid-K spectral type. In these cases the weak photospheric features from their spectra make their spectral types poorly constrained.

\begin{figure*}
\begin{center}
\includegraphics[angle=0,width=1\columnwidth]{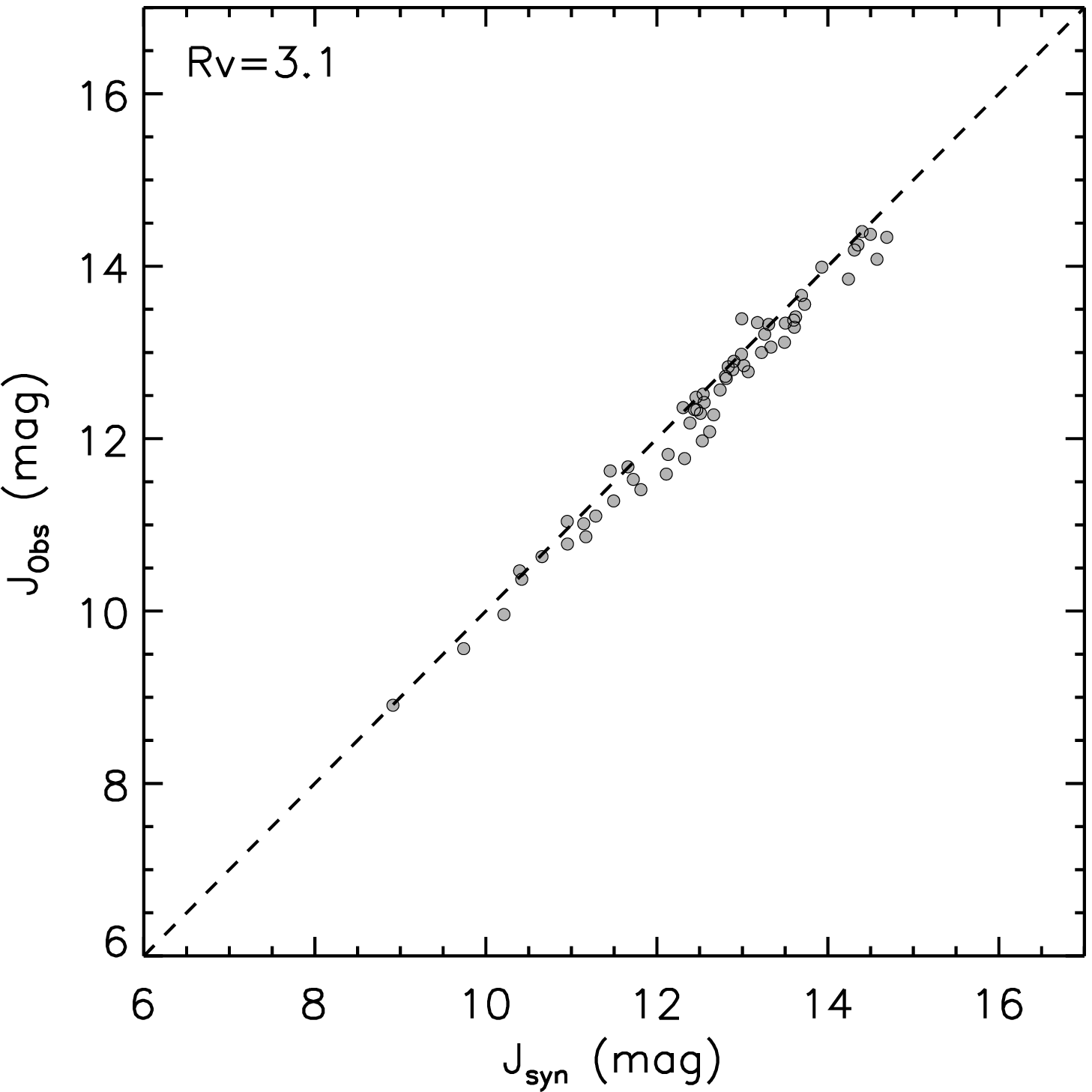}
\includegraphics[angle=0,width=1\columnwidth]{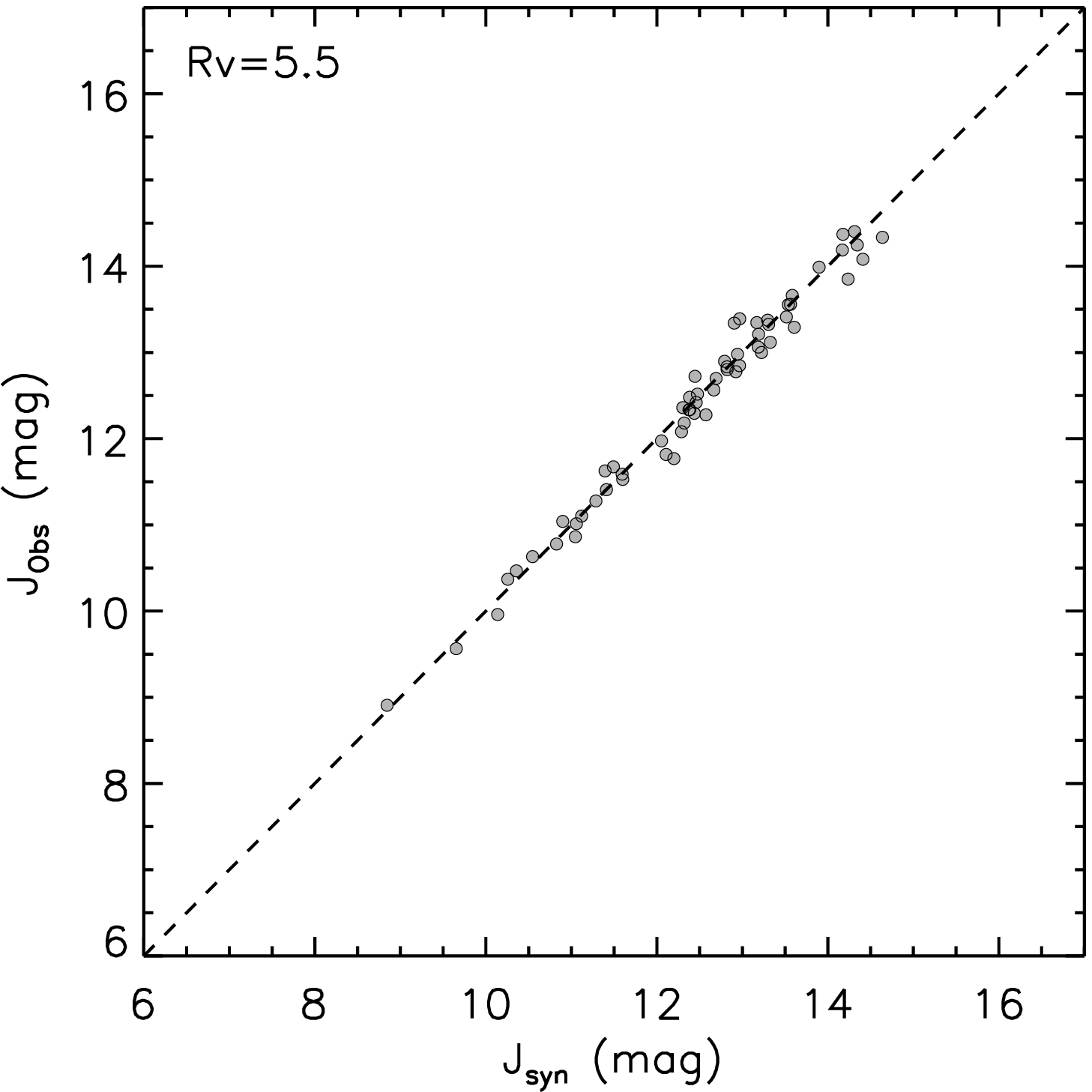}
\caption{Comparison of the observed 2MASS J-band photometry and synthetic 2MASS J-band photometry derived from the best-fit PMS spectral templates  for the diskless stars in the field. The dashed line in each panel shows the 1:1 relation.} \label{Fig:Jmag}
\end{center}
\end{figure*}

\begin{figure}
\begin{center}
\includegraphics[angle=0,width=1\columnwidth]{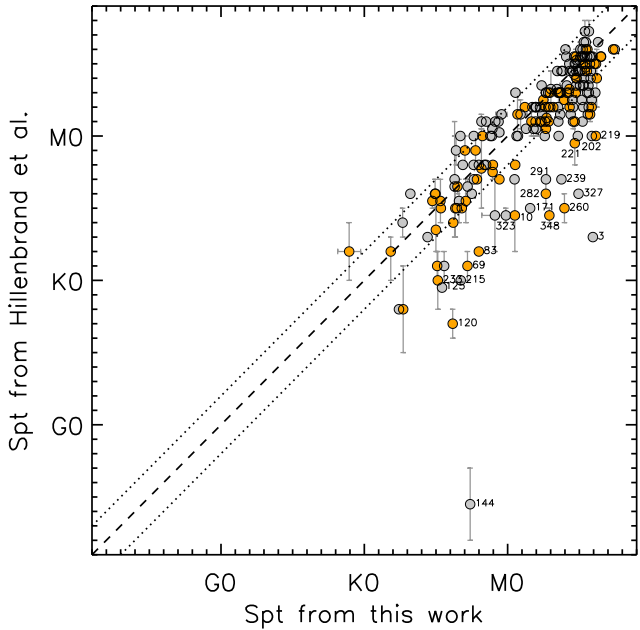}
 \caption{Comparison of the spectral types from the literature \citep{1997AJ....113.1733H,2013AJ....146...85H} and from this work with $R_{\rm V}$=5.5 and constant Accretion Continuum Spectrum.  The dashed line in each panel shows the 1:1 relation, and the dotted lines indicate the two subclass differences from the dashed line. The gray-color filled circles are for the sources with  $r_{\rm ex,\, 7465}<0.2$ and the orange-color filled circles are for those with  $r_{\rm ex,\, 7465}\geq0.2$. The sources with spectral-type difference larger than five subclasses are marked with their identification numbers in Table~\ref{tab:tabe_ONC}.\label{Fig:SPT_COM}}

\end{center}
\end{figure}

\begin{figure*}
\begin{center}
\includegraphics[angle=0,width=2\columnwidth]{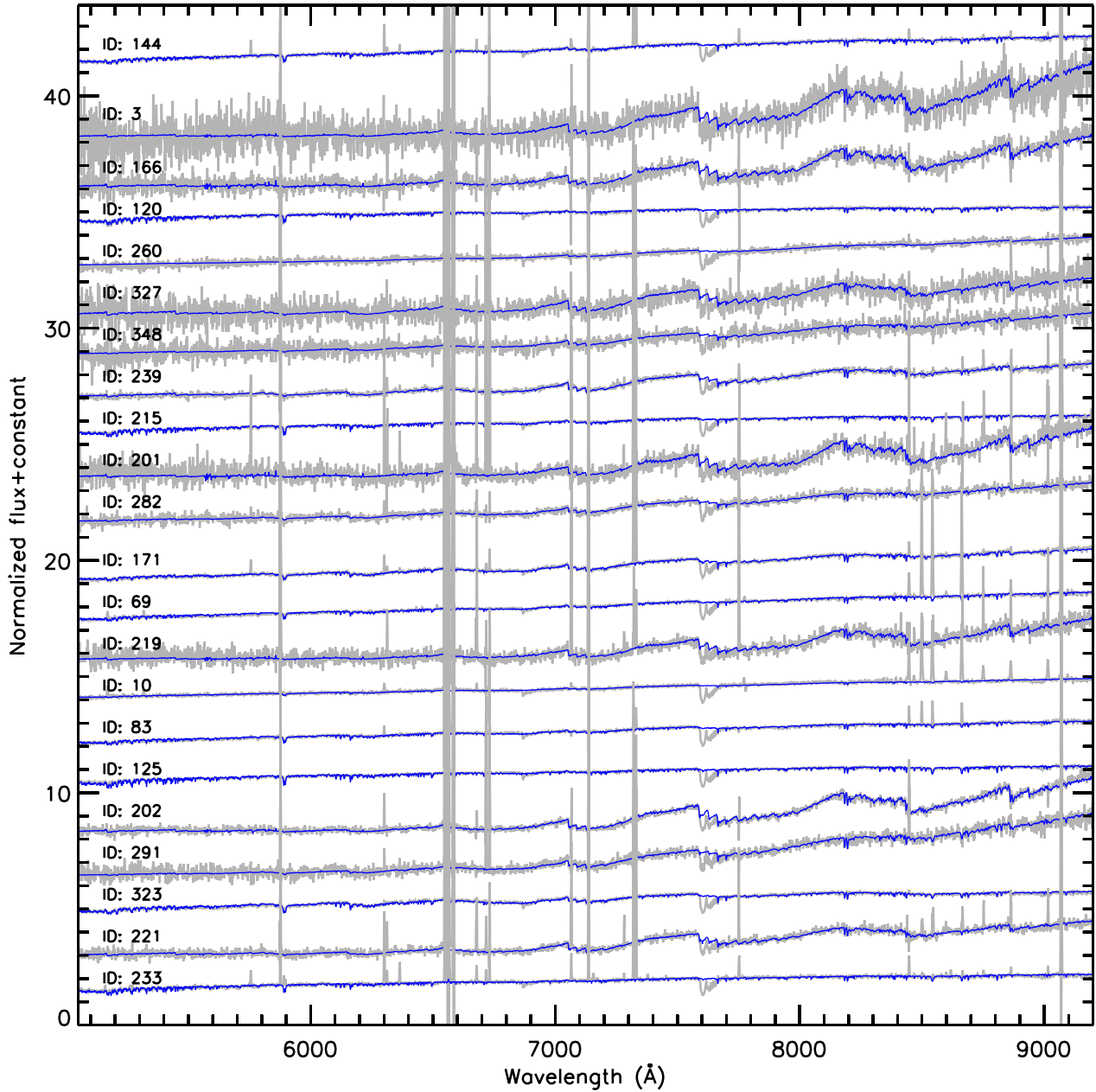}
\caption{Best-fit spectral templates (blue) over-plotted on the MUSE spectra (gray) of {\newrev 22} sources which have spectral type differences larger than five subclasses between the ones from \citet{2013AJ....146...85H} and this work. {\rev From top to bottom, the spectral type difference decreases}, and the identification number of each source is the same as in Table~\ref{tab:tabe_ONC}. }
\label{Fig:SPT_dif}
\end{center}
\end{figure*}

\subsection{Stellar properties}
We convert the spectral types to  effective temperatures (\Teff) using the conversions in \cite{2017AJ....153..188F}, which are from \cite{2013ApJS..208....9P} for stars earlier than M4 and from \cite{2014ApJ...786...97H} for stars later than M4 type. We obtain the stellar luminosity (\Lstar) using the \Teff\ and the photospheric flux at 7510~\AA\ from the best-fit templates described above and the Bolometric Corrections from \cite{2014ApJ...786...97H}.  A distance of 388$\pm$10~pc {\newnewrev measured from the multi-epochs observations with the Very Large Baseline Array \citep{2017ApJ...834..142K}, confirmed by the {\it GAIA} measurement (389$\pm$3~pc, \citealt{2018AJ....156...84K})}, is used for computing the bolometric luminosity \Lstar.

Among the different  $R_{\rm V}$ values and the shapes of Accretion Continuum Spectrum, there are systematic differences on the derived \Lstar. Based on these comparisons we find that (1) for a given $R_{\rm V}$, the \Lstar\ obtained under the assumption of constant Accretion Continuum Spectrum is systematically lower by $\sim$0.07~dex compared to the \Lstar\ assuming the Accretion Continuum Spectrum as a black body emission with $T=7000$~K; (2) for a given shape of the Accretion Continuum Spectrum, the stellar luminosity from $R_{\rm V}=3.1$ is systematically lower by $\sim$0.1~dex than that assuming $R_{\rm V}=5.5$.

In Table~\ref{tab:tabe_ONC}, we list the spectral types, \Lstar\, $A_{V}$, and veiling $r_{7465}$ for different values of $R_{V}$ and shapes of Accretion Continuum Spectrum for all the sources discussed in this work.

\begin{figure*}
\begin{center}
\includegraphics[angle=0,width=2\columnwidth]{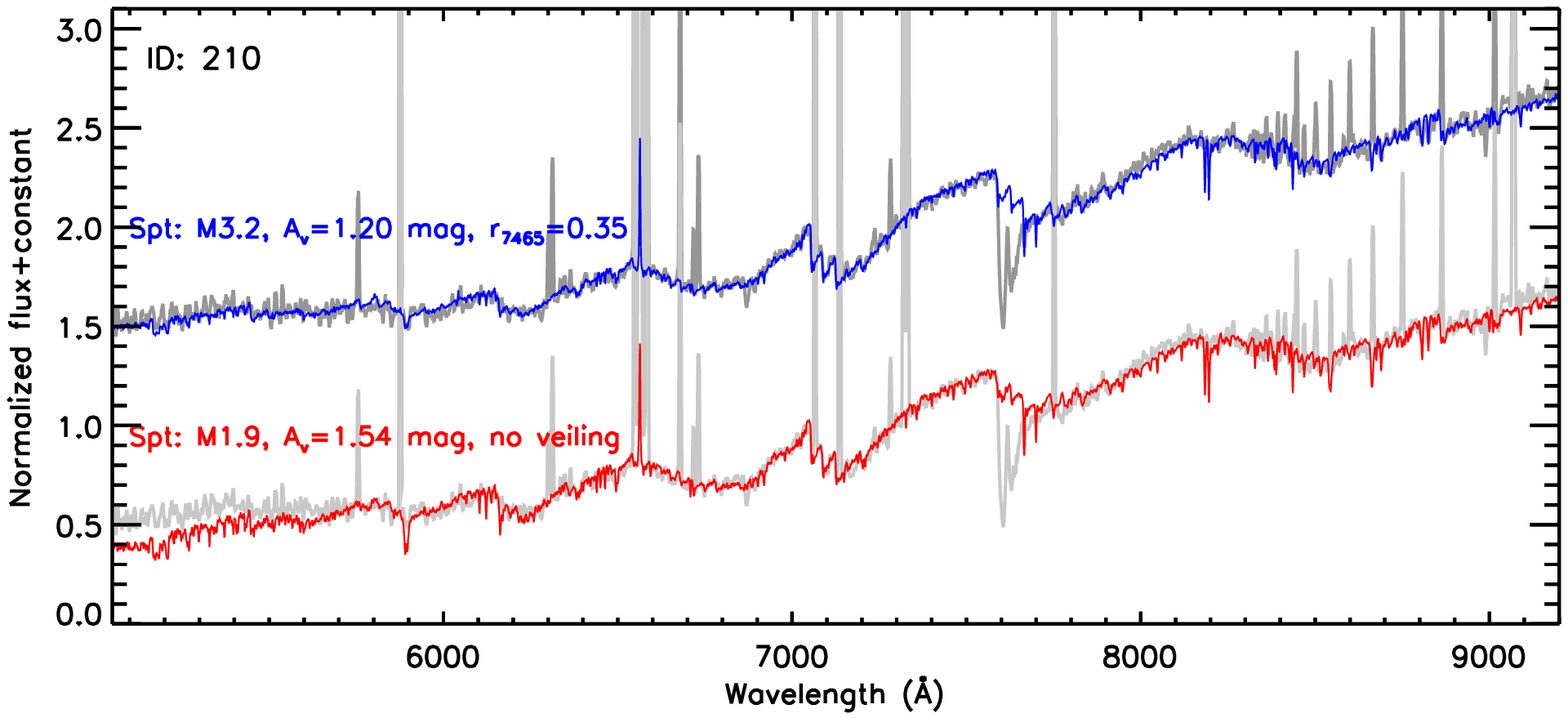}
\caption{Best-fit spectral templates over-plotted on the MUSE spectra (gray) of  Source 210. The best-fit templates are obtained in two cases: with constant Accretion Continuum Spectrum (blue) and without veiling (red), assuming $R_{V}=5.5$.}
\label{Fig:SPT_veil}
\end{center}
\end{figure*}

\section{Extinction law}\label{Sect:extlaw}
In this work, we use the extinction law from \citet{1989ApJ...345..245C}, and take $R_{V}$=3.1 and 5.5. As discussed in Appendix~\ref{Appen:cali}, the value of $R_{V}$=5.5 can fit the spectral energy distribution of HD~37042 in near-infrared bands better than $R_{V}$=3.1. $R_{V}$=5.5 has been considered more appropriate for the Orion nebula cluster  \citep{1965ApJ...141..923J,1968ApJ...152..913L,1970BOTT....5..229C,2010A&A...518A..62G}. 
We further test if $R_{V}$=5.5 is appropriate for our targets by comparing the observed 2MASS $J$-band photometry and synthetic 2MASS $J$-band photometry derived from the best-fit PMS spectral templates for the diskless stars in the field; see Fig.~\ref{Fig:Jmag}. 
We select diskless stars which show no excess emission in mid-infrared bands \citep{2001ApJ...558L..51M,2012AJ....144..192M,2005AJ....129.1534R,2005AJ....130.1763S}.  

The comparisons show that the synthetic photometry derived from the best-fit PMS templates with $R_{V}$=5.5 agrees well with the observed photometry with the mean difference of less than 1\%, while the comparisons with $R_{V}$=3.1 underestimate the flux in $J$~band by $\sim$15\%. This indicates that $R_{V}$=5.5 is a better choice for our targets.  In the following sections, we will include two $R_{V}$ values  and two types of the Accretion Continuum Spectrum only in \sect~\ref{Sect:HRD} and also in Table~\ref{tab:tabe_ONC} in order to show how the different $R_{V}$ affects the results in determining the stellar parameters, and adopt $R_V$=5.5 and constant Accretion Continuum Spectrum in other sections.

\section{Comparison with literature}\label{SECT:Comparison}
\subsection{Spectral type}
\cite{1997AJ....113.1733H} carried out the first extensive spectroscopic survey of young stars in the Orion nebula cluster,  which are further enhanced by \cite{2013AJ....146...85H}. Figure~\ref{Fig:SPT_COM} compares the spectral types of 203 common sources from the literature \citep{1997AJ....113.1733H,2013AJ....146...85H} and this study.
For the sources with the spectral types from both \cite{1997AJ....113.1733H} 
and \cite{2013AJ....146...85H}, we 
only show the spectral types from the latter one in Figure~\ref{Fig:SPT_COM}.  Among these common sources, 40\% of them show spectral-type differences of less than one subclass, and $\sim$68\% with spectral-type differences of less than two subclasses. There are {\newrev 22} sources with differences larger than five subclasses. In particular,  Source 144 has a spectral type of F2-F7 in \citet{1997AJ....113.1733H}, a spectral type of K0$-$K7 in \cite{2000ApJ...540.1016L}, and a spectral type of K7 in this work. In Fig.~\ref{Fig:SPT_dif}, we over-plot the best-fit spectral template on the MUSE spectrum of Source 144. We find that the template fits the observed spectrum very well. In the figure, we also show the 21 other sources with spectral type difference larger than five subclasses. For each source, the best-fit template can reproduce the observation reasonably well.

\begin{figure*}
\begin{center}
\includegraphics[angle=0,width=1\columnwidth]{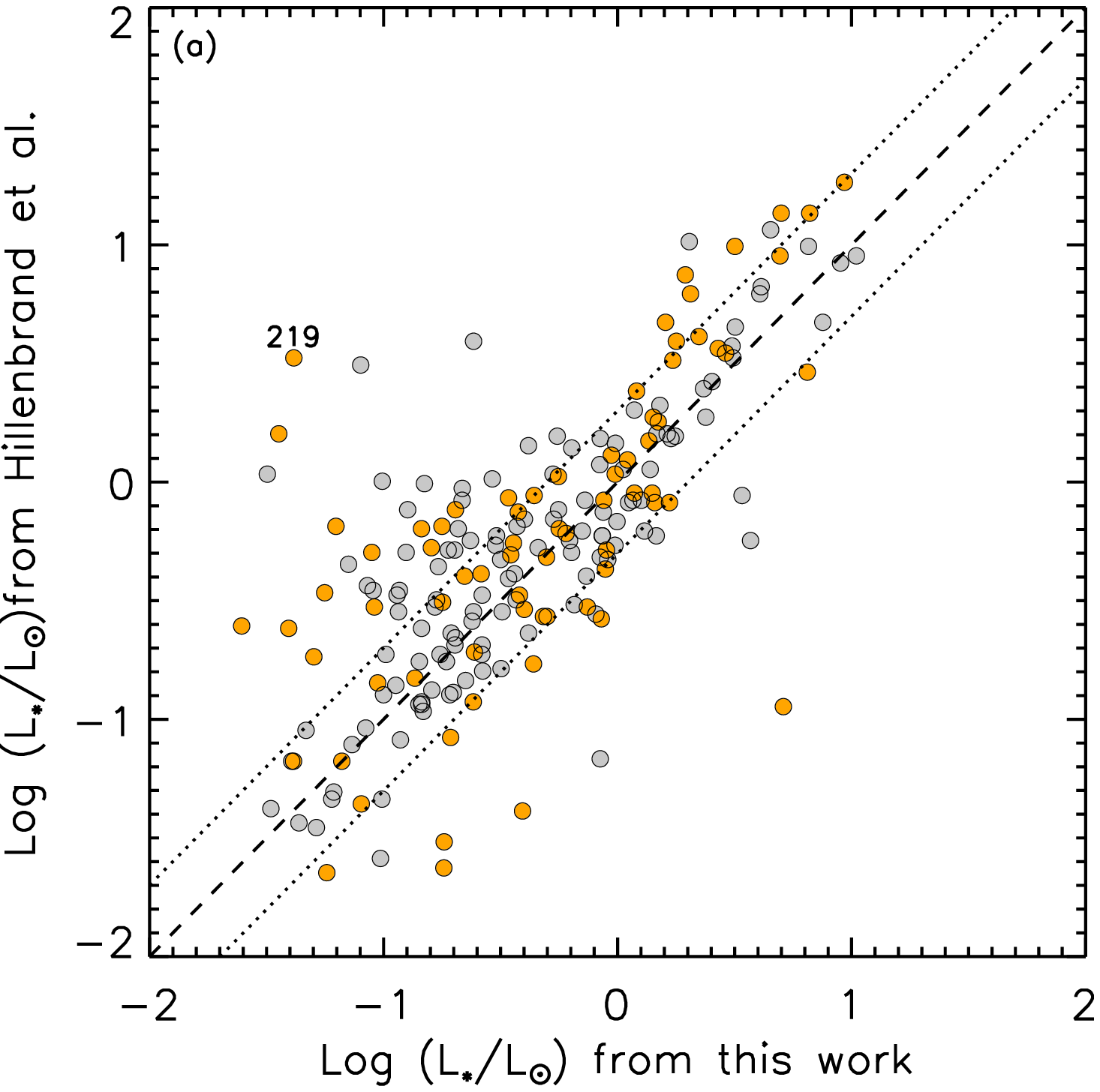}
\includegraphics[angle=0,width=1\columnwidth]{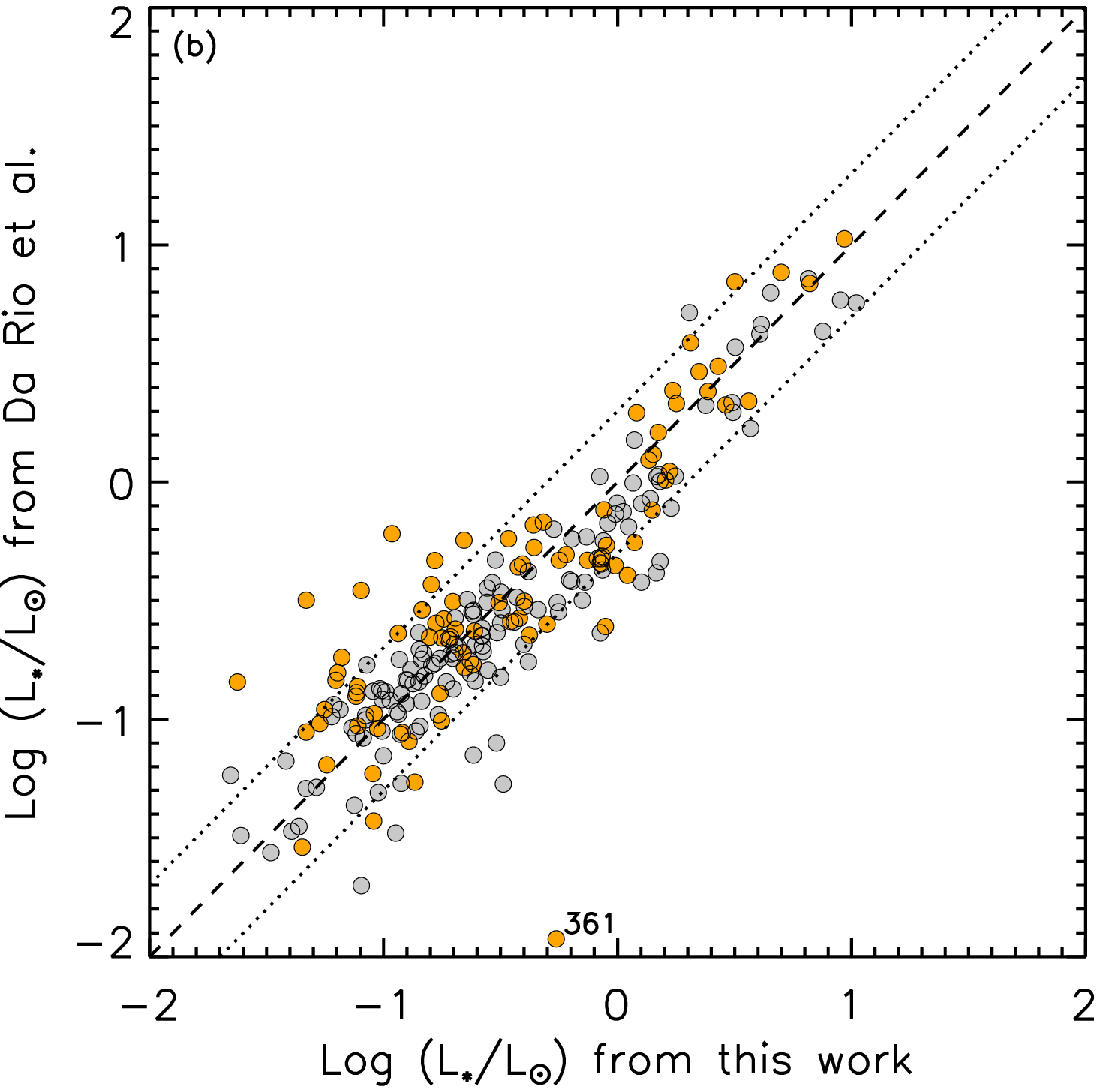}
\caption{Comparison of the stellar luminosity from this work with those from \cite{1997AJ....113.1733H} (a)  and from \cite{2012ApJ...748...14D} (b). In each panel, the dashed line shows the 1:1 relation, and the dotted lines indicate the difference in the stellar luminosity  by a factor of 2. The gray-color filled circles are for sources with  $r_{7465}<0.2$ and the orange-color filled circles for those with  $r_{7465}\geq0.2$. The stellar luminosity from this work are derived taking $R_{V}$=5.5 and assuming a constant Accretion Continuum Spectrum. 
}
\label{Fig:AV_Lum_com}
\end{center}
\end{figure*}

Beside the extreme cases discussed above, literature spectral types for a number of sources in the cluster are actually very uncertain. In Table~\ref{tab:tabe_ONC}, we list all the known spectral types of individual sources collected from \cite{2013AJ....146...85H}. There can be 2$-$3 subclass differences between the literature values and our work. The differences are mostly due to (1) differences in the spectral ranges used for spectral typing and (2)  whether veiling has been considered. For an accreting source, the Accretion Continuum Spectrum affects the blue part of the spectrum much more significantly than the red part. Therefore,  fitting just the blue part of the spectrum  results in earlier spectral type than using the red part, if  veiling is not considered. 
For example, Figure~\ref{Fig:SPT_veil} shows the effect of veiling for 
Source 210.  Not accounting for veiling leads to a M1.9 spectral type, very similar to the M2 spectral type reported in \cite{1997AJ....113.1733H}. When veiling is considered, however, the derived spectral type is one subclass later. It is also clear that the best-fit template with veiling reproduces the observations much better at wavelengths $\lesssim$ 6200\,\AA\ than that the template without veiling. However, the difference between the two best-fit templates is minor at longer wavelengths, illustrating the importance of broad wavelength range for accurate spectral classification. A similar example can be found in Figure~15 in \cite{2014ApJ...786...97H}.

\subsection{Stellar luminosity}

Figure~\ref{Fig:AV_Lum_com}(a) compares $L_{\star}$ from this work with those in \cite{1997AJ....113.1733H}. There are 199 common sources between the two works.  The $L_{\star}$  from this work are derived using $R_{V}$=5.5 and assuming a constant Accretion Continuum Spectrum. The $L_{\star}$ from \cite{1997AJ....113.1733H} have been scaled from 470~pc to the new distance of 388~pc \citep{2017ApJ...834..142K} which is also used in our work.
The mean difference on the $L_{\star}$ between two works  is  0.26~dex. There are 37\% of common sources with difference larger than a factor of two. The largest difference comes from Source 219 with  Log~$(L_{\star}/L_{\odot})$=0.52 from \cite{1997AJ....113.1733H}  and Log~$(L_{\star}/L_{\odot})$=$-$1.38 from this work. This object also has very different spectral types between both works: M0 from \cite{1997AJ....113.1733H}  and M6 from this work; see Fig.~\ref{Fig:SPT_dif}. The synthetic photometry of our best-fit template for this source is  13.74~mag in $J$ band, which is fainter than the observed one 13.20--13.47~mag \cite{2010AJ....139..950R}. This difference can be due to infrared excess emission from the circumstellar disk, which is suggested by its infrared color ($K-L$=0.82 mag from \citealt{2002ApJ...573..366M}) and strong optical veiling (see Table~\ref{tab:tabe_ONC}). 

\begin{figure*}
\begin{center}
\includegraphics[angle=0,width=1\columnwidth]{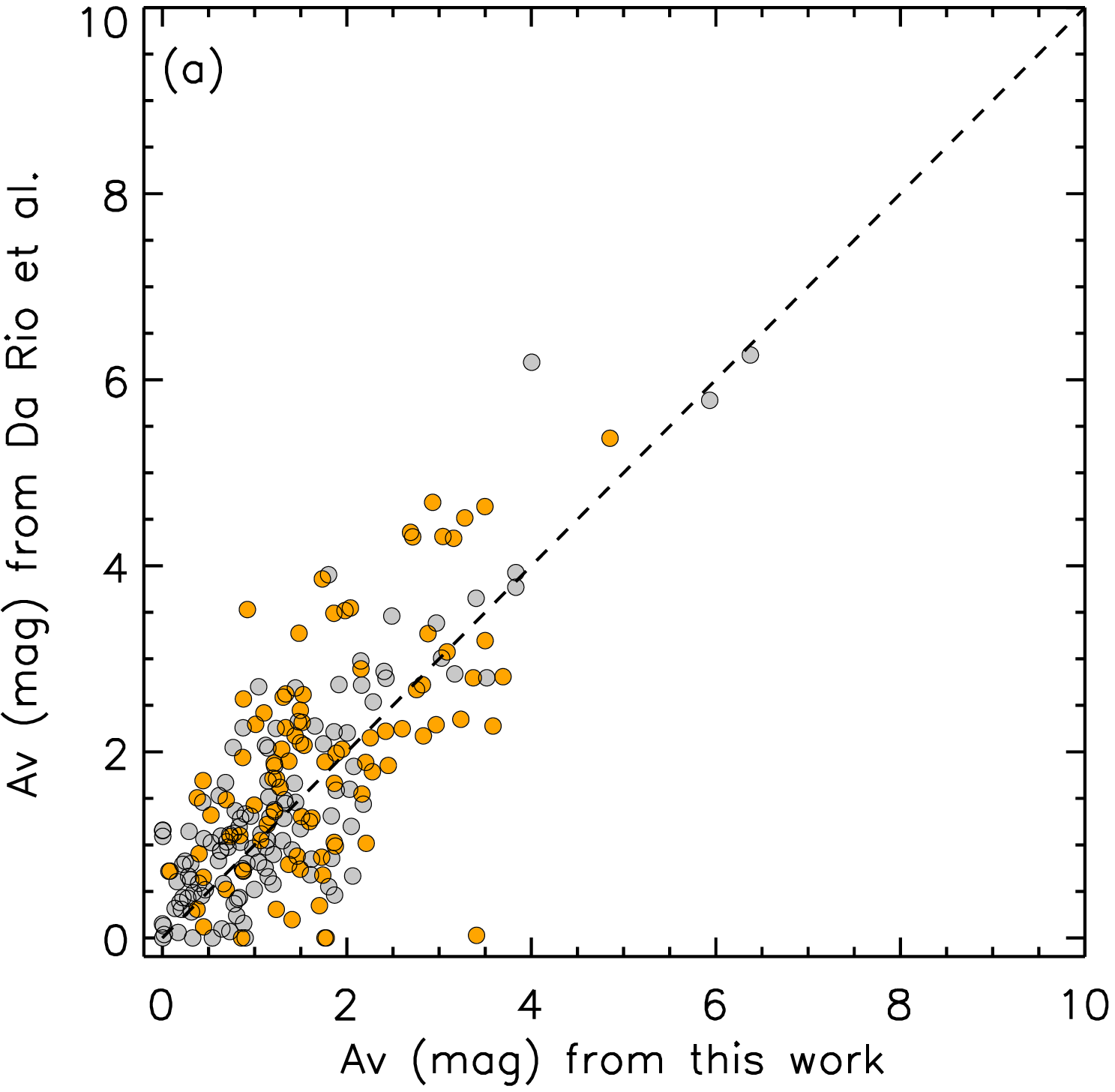}
\includegraphics[angle=0,width=1\columnwidth]{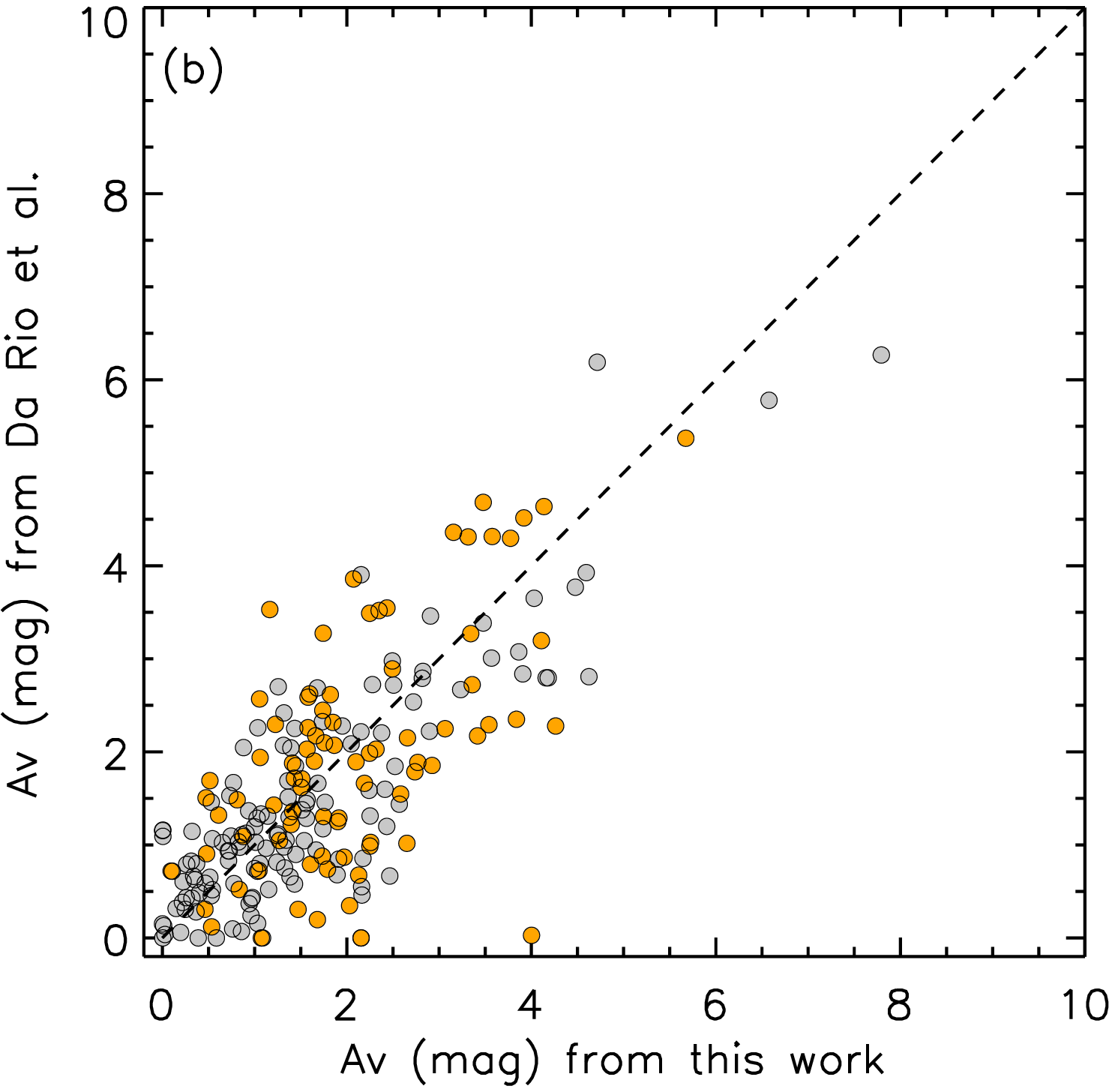}
\includegraphics[angle=0,width=1\columnwidth]{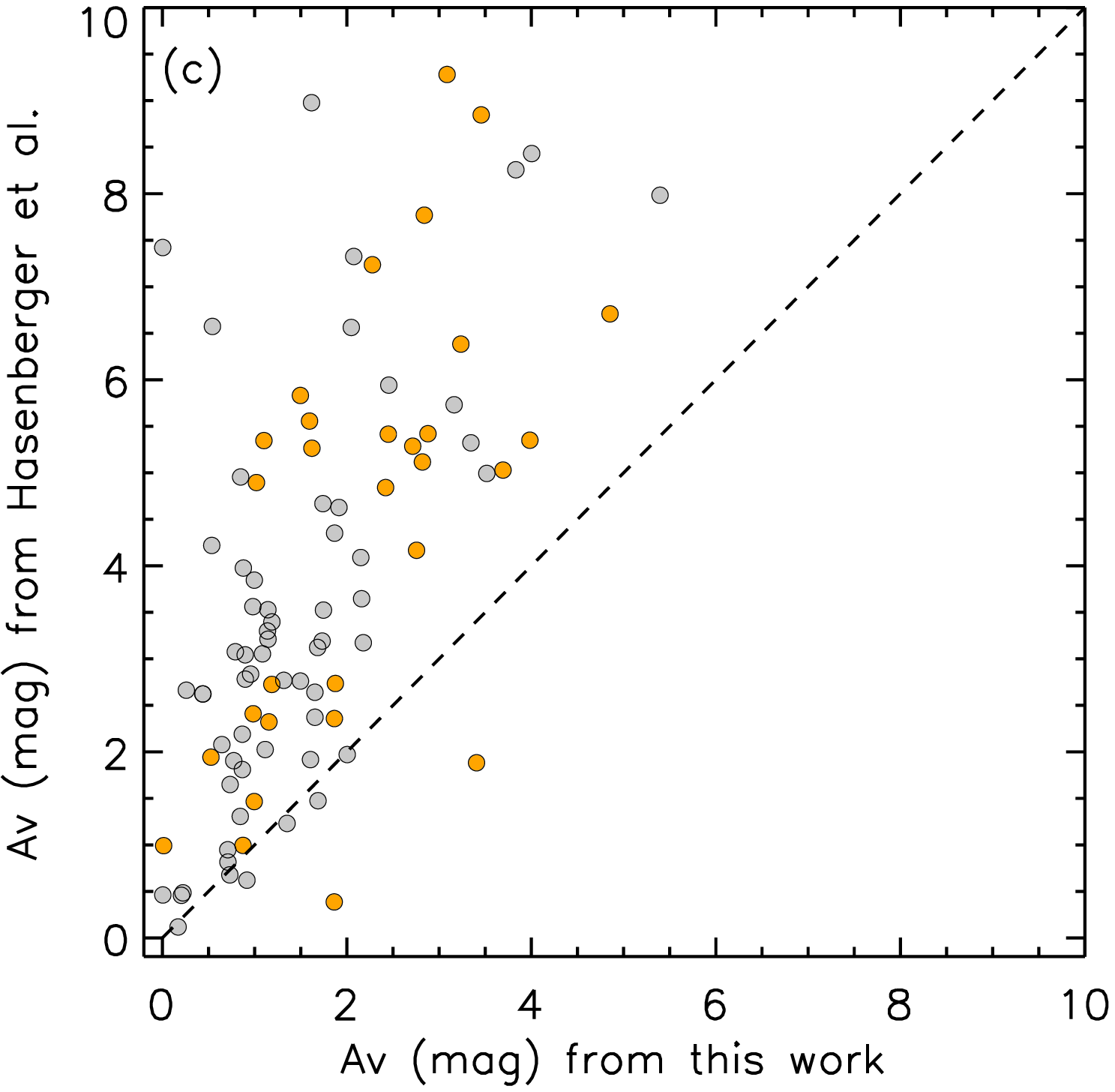}
\includegraphics[angle=0,width=1\columnwidth]{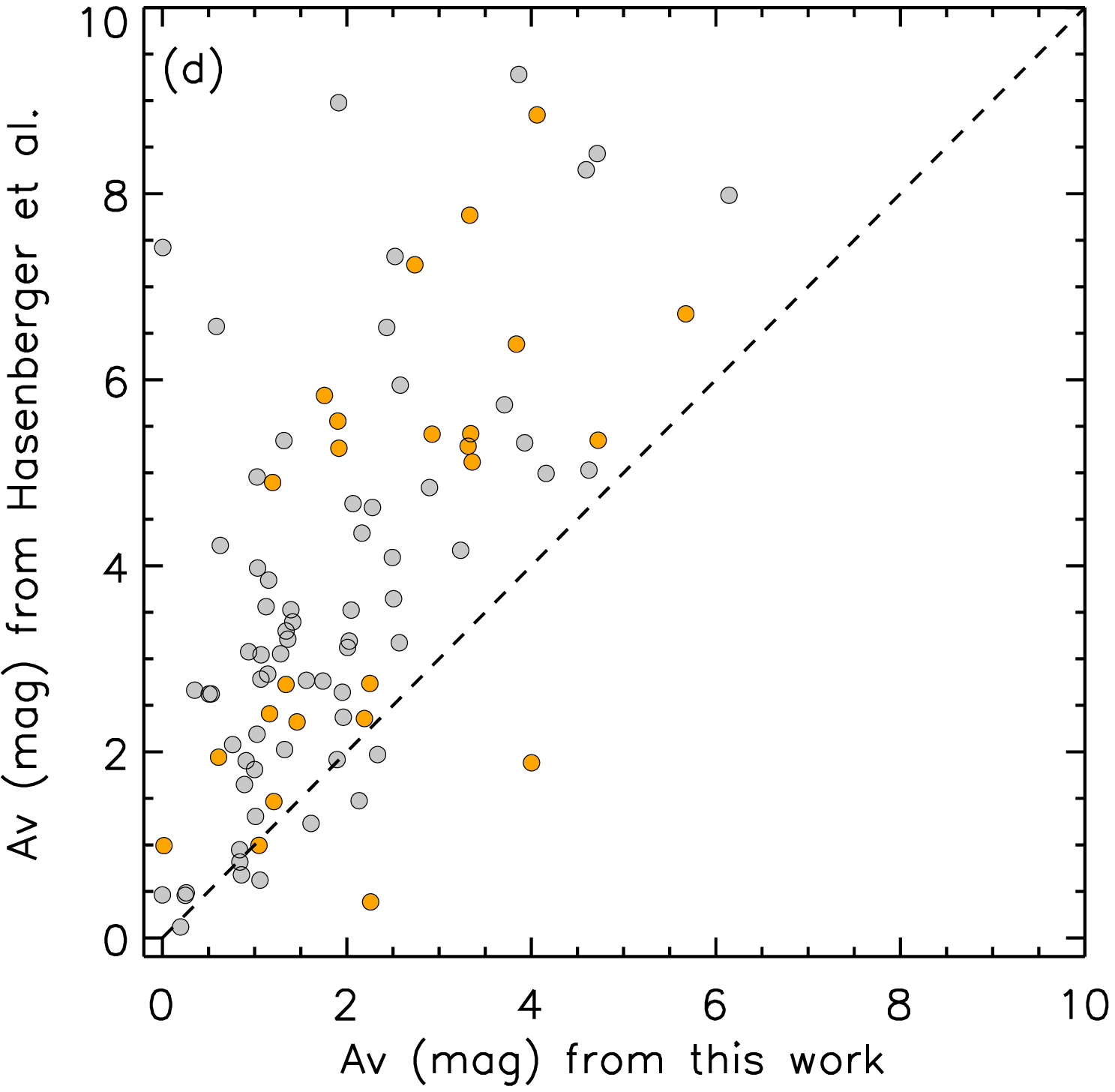}
\caption{{\newrev Comparisons of the extinction from this work with those from \cite{2012ApJ...748...14D} (a, b) and \cite{2016A&A...593A...7H} (c, d). The extinction from this work are derived assuming a constant Accretion Continuum Spectrum and taking $R_{V}$=3.1 in Panel~(a, c) and $R_{V}$=5.5 in Panel~(b, d).  In each panel, the dashed line shows the 1:1 relation. The gray-color filled circles are for sources with  $r_{7465}<0.2$ and the orange-color filled circles for those with  $r_{7465}\geq0.2$. }}
\label{Fig:AV_com}
\end{center}
\end{figure*}

\cite{2012ApJ...748...14D} {\newrev complemented the sources with the spectral types from \cite{1997AJ....113.1733H} by utilizing the photometry in $I$ band and two medium-band filters at at $\lambda\sim$7530 and 7700~\AA\ to derive the spectral types for M-type stars.} In Figure~\ref{Fig:AV_Lum_com}(b) we compare $L_{\star}$ from this work with the ones from \cite{2012ApJ...748...14D} for the {\newrev 227} sources in common. The $L_{\star}$  from our work are same as above, and the $L_{\star}$ from \cite{2012ApJ...748...14D} have been scaled to the distance we adopted (388~pc). The mean difference of the $L_{\star}$ between two works is 0.16~dex, and only 16\% of the common sources have differences larger than a factor of two. Among them, source 361 shows the largest difference: Log~($L_{\star}/L_{\odot}$)=$-$1.897 from  \cite{2012ApJ...748...14D} and Log~($L_{\star}/L_{\odot}$)=$-$0.262 from our work. This is mainly due to the difference of the spectral types used in  the two analyses: G4--K5 used in \cite{2012ApJ...748...14D} which was taken from \cite{2000ApJ...540.1016L} and K8.3 in our work. The difference is mainly due to its strong veiling ($r_{7465}$) with large infrared excess emission ($K-L$=1.33 mag from \citealt{2002ApJ...573..366M}). 

 Both \cite{1997AJ....113.1733H} and  \cite{2012ApJ...748...14D} take $R_{V}$=3.1. If we take the same $R_{V}$=3.1 and assume a constant Accretion Continuum Spectrum, the mean difference on $L_{\star}$ from our work and \cite{1997AJ....113.1733H} becomes even larger (0.31~dex) with 46\% of common sources with difference larger than a factor of two. However, the mean difference on $L_{\star}$ from our work and \cite{2012ApJ...748...14D} becomes smaller (0.14~dex) with 18\% of common sources with difference larger than a factor of two.   
 
 \begin{figure*}
\begin{center}
\includegraphics[angle=0,width=1\columnwidth]{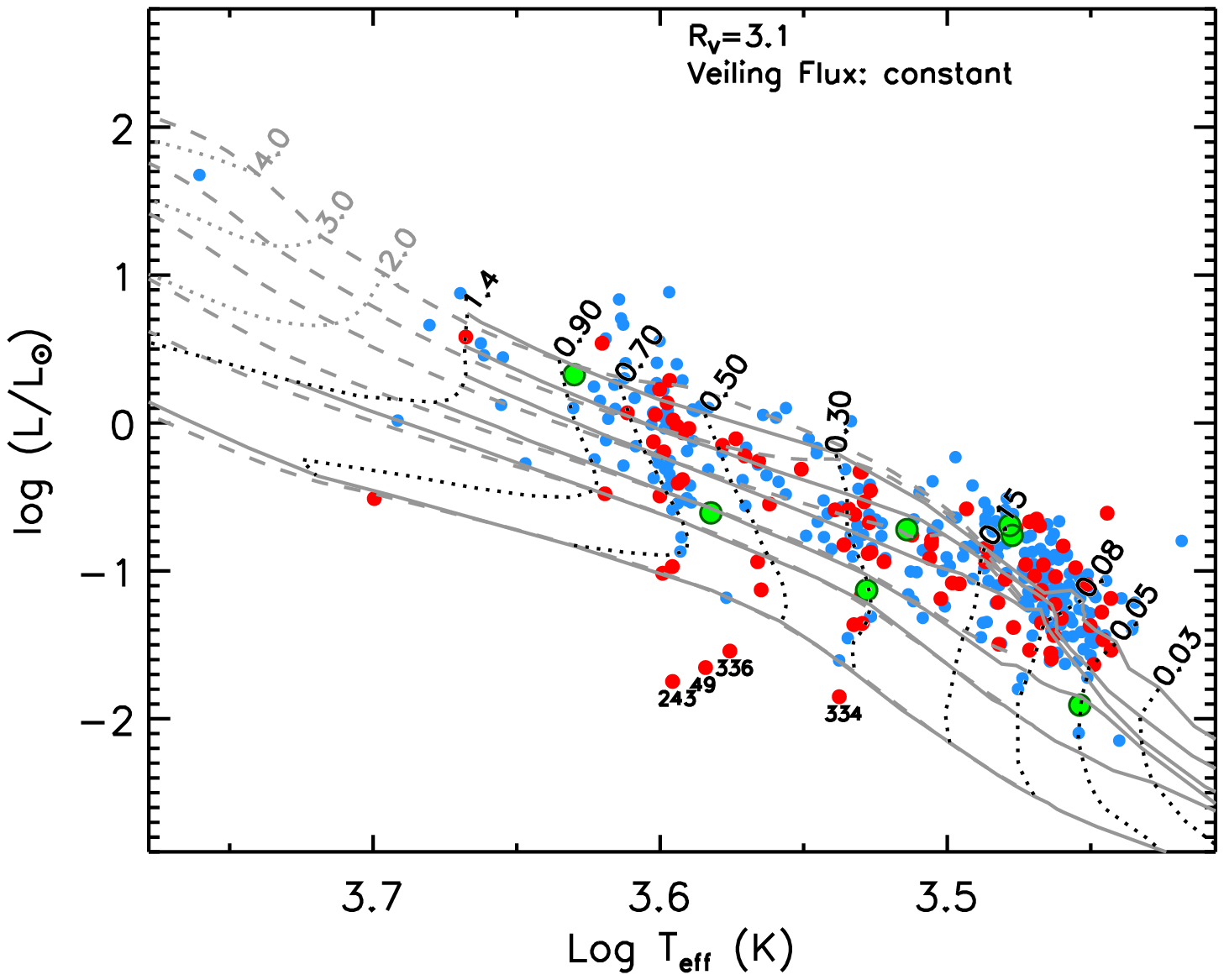}
\includegraphics[angle=0,width=1\columnwidth]{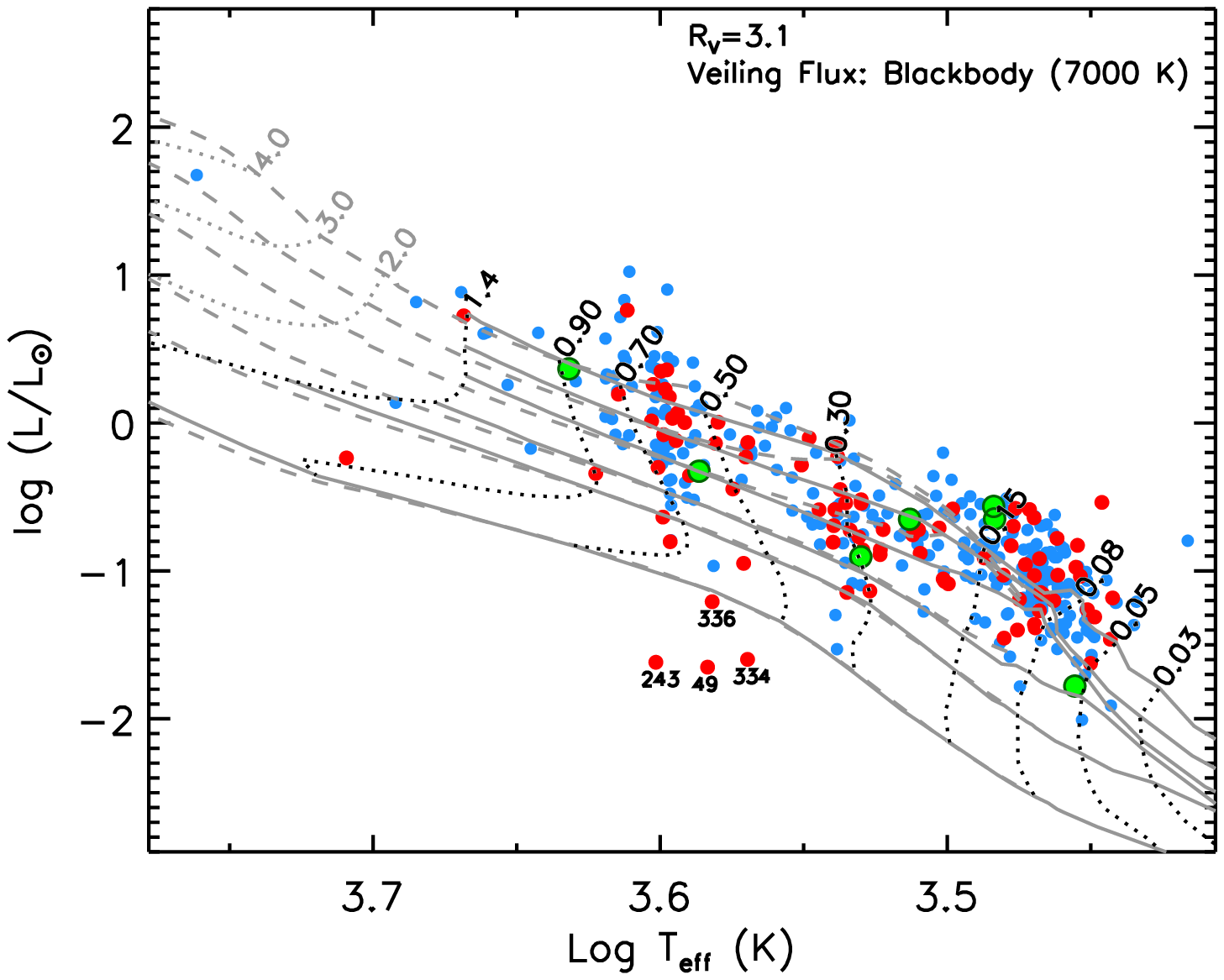}
\includegraphics[angle=0,width=1\columnwidth]{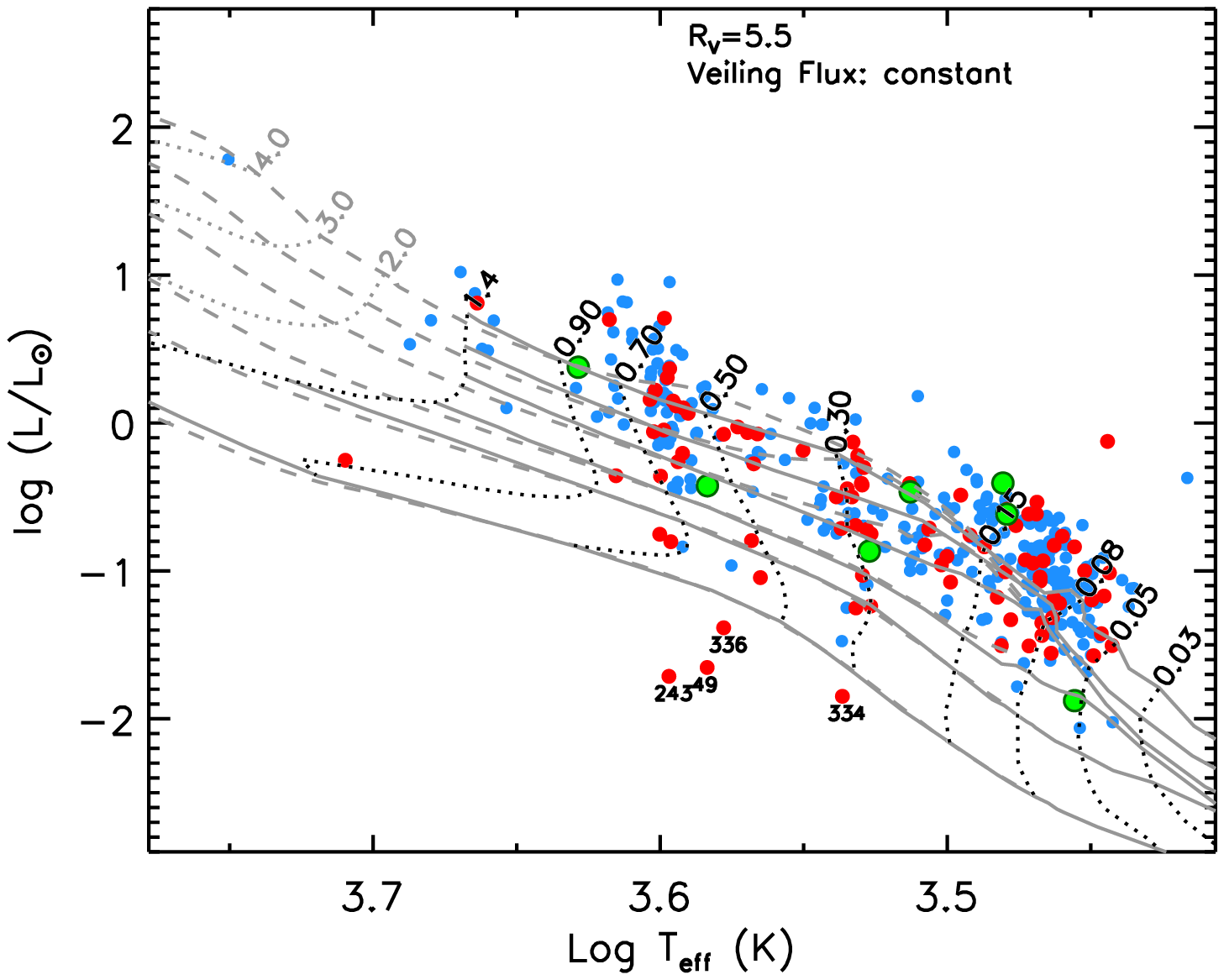}
\includegraphics[angle=0,width=1\columnwidth]{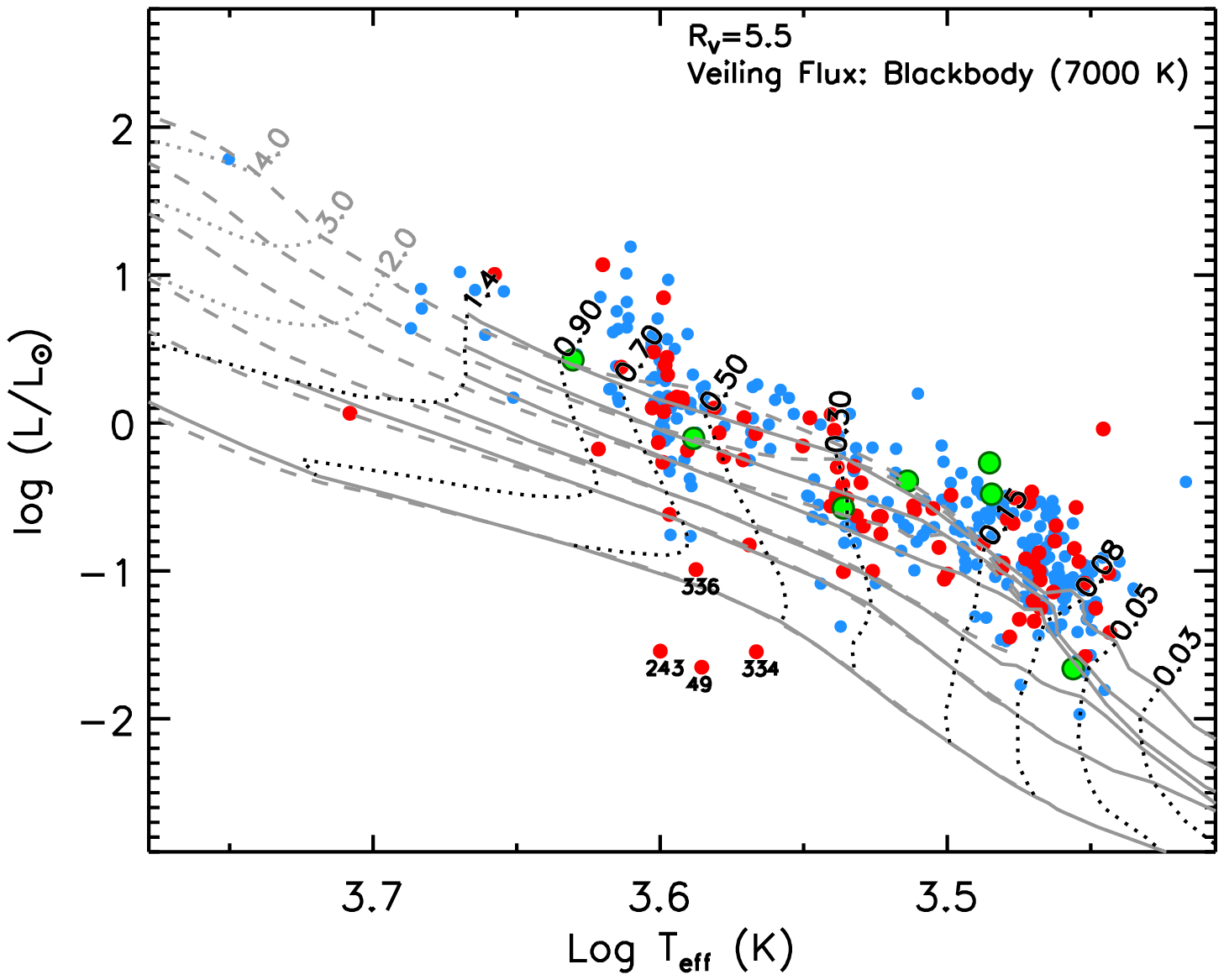}
\caption{HR diagrams for the sources in the Trapezium cluster. The stellar parameters are derived with different values of $R_{\rm V}$ and  shapes of Accretion Continuum Spectrum and shown in different panels.
In each panel, the red-color filled circles are sources with proplyds, green-color filled circles are ones with silhouette disks, and blue-color filled circles are other sources in our sample.  Four subluminous proplyds are marked with their identification numbers in Table~\ref{tab:tabe_ONC}.
The nonmagnetic evolutionary tracks (dashed lines) from \cite{2016A&A...593A..99F} and the evolutionary tracks (solid lines) from \cite{2015A&A...577A..42B} are shown as a comparison. The isochrones are at ages of 0.5, 1, 2,  5, 10, and 50~Myr from top to bottom. 
}
\label{Fig:HRD}
\end{center}
\end{figure*}

 \subsection{Extinction}
{\newrev 
  In Figure~\ref{Fig:AV_com} (a, b), we compare the $A_{\rm V}$ from this work with those in \cite{2012ApJ...748...14D}.  In these figures, the $A_{\rm V}$ from this work are derived assuming a constant Accretion Continuum Spectrum  and taking $R_{V}$=3.1 in Figure~\ref{Fig:AV_com}(a)  and  $R_{V}$=5.5 in Figure~\ref{Fig:AV_com}(b). In \cite{2012ApJ...748...14D} they adopted the extinction law from \cite{1989ApJ...345..245C} with  $R_{V}$=3.1 and derived the extinction in optical bands. 
  In Figure~\ref{Fig:AV_com}(a),  the mean difference between the two works is 0.18~mag with the standard deviation of 0.81~mag, and in Figure~\ref{Fig:AV_com}(b) these are $-$0.10~mag and 0.87~mag, respectively. The main sources of difference in extinction are (1) the differences in the assigned SpT, (2) whether one includes veiling in deriving the extinction, (3) whether the photometry of the sources are obtained simultaneously. In general, the $A_{\rm V}$  differences found here agree well with those discussed in the literature (see e.g. \citealt{2014ApJ...786...97H}).

  Figure~\ref{Fig:AV_com} (c, d) compare the  $A_{\rm V}$ from this work with those from \cite{2016A&A...593A...7H}. The extinction in \cite{2016A&A...593A...7H} are derived using the photometry in near-infrared $J, H, K_{\rm s}$ bands and spectral types from \cite{2013AJ....146...85H}, adopting the extinction law from \cite{1989ApJ...345..245C} with $R_{V}$=3.1. The $A_{\rm V}$ measured from the photometry in near-infrared bands are significantly higher than those derived at optical wavelengths. The large differences on the extinction derived at optical wavelengths and near-infrared wavelengths have been noted in literature \citep{2014ApJ...780..150M,2014ApJ...786...97H}, and could be caused by the large near-infrared excess emitted from the inner regions of circumstellar disks \citep{2014ApJ...786...97H}. }

\begin{figure}
\begin{center}
\includegraphics[angle=0,width=1\columnwidth]{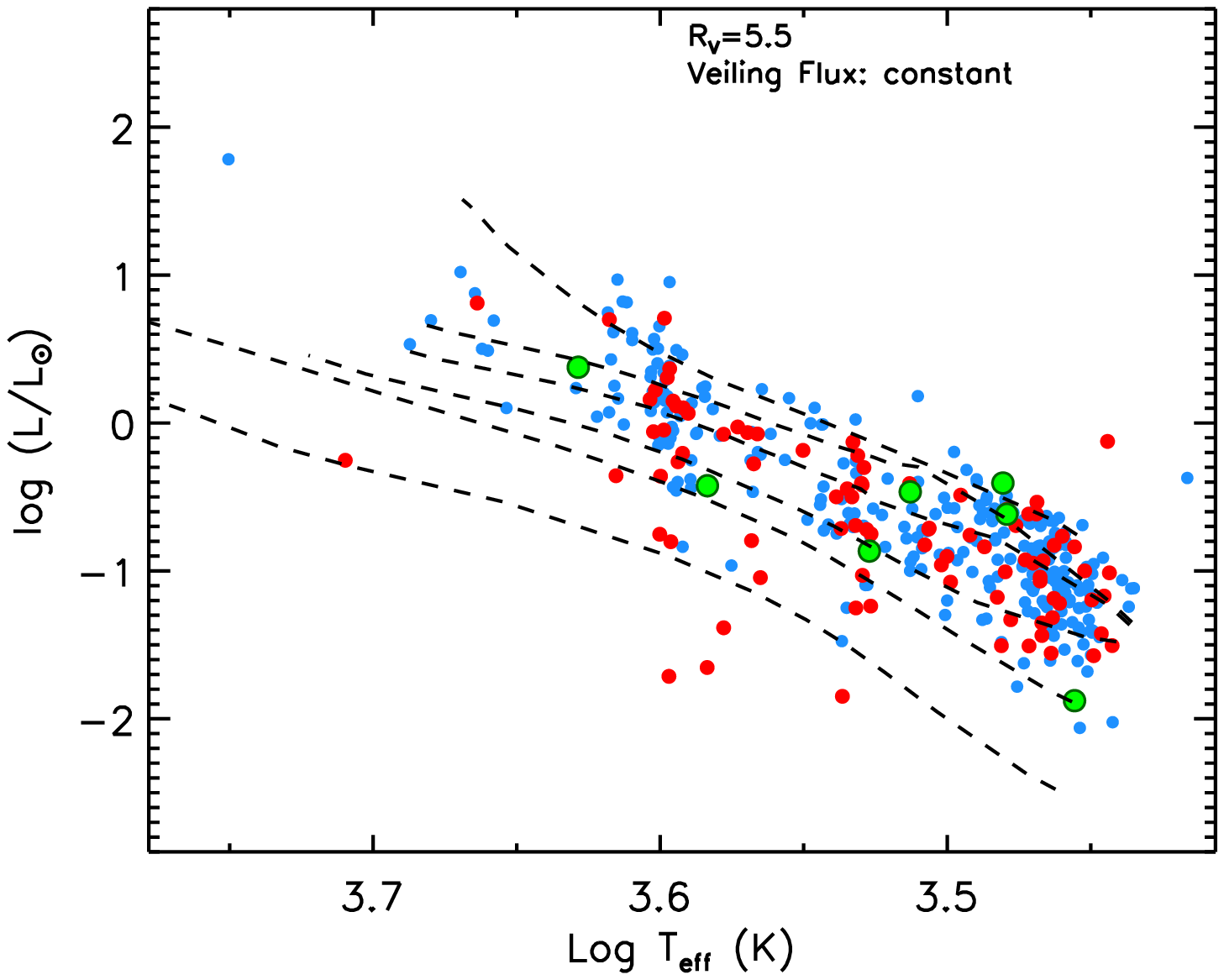}
\caption{HR diagram for the sources in Trapezium cluster. The PMS evolutionary tracks are from \cite{2016A&A...593A..99F}, in which the influence by the presence of magnetic fields are included in the calculation. The black dash lines show  isochrones for  0.1, 1, 2, 5, 10, and 50~Myr, from top to bottom.}
\label{Fig:magHRD}
\end{center}
\end{figure}

\section{Herzsprung-Russell diagram}\label{Sect:HRD}

With the derived \Teff\ and bolometric luminosities for different values of $R_{V}$ and shapes of Accretion Continuum Spectrum, we place the stars in the H-R diagram in Figure~\ref{Fig:HRD}. As a comparison, we show the nonmagnetic  evolutionary tracks  from \cite{2016A&A...593A..99F} and  from \cite{2015A&A...577A..42B}.  A combination of the two evolutionary tracks has been used in \cite{2016ApJ...831..125P} and can match empirical stellar loci of nearby young associations in H-R diagram  better than older models \citep{2015ApJ...808...23H}.  We match our sample with those in \cite{2008AJ....136.2136R}, and find 89 sources with externally ionized protoplanetary disks (proplyds), as well as 7 sources with disks seen only in absorption against the bright nebular background (silhouette disks). In Figure~\ref{Fig:HRD}, all these sources are shown  with different symbols. There are a few proplyds, which appear subluminous compared to the sources of the same spectral types. We will discuss these subluminous sources in Section~\ref{sect:sub_proplyd}

In Figure~\ref{Fig:HRD}, depending on the adopted $R_{\rm V}$ and the shapes of Accretion Continuum Spectrum,  there are 73\%--89\% of the sources above the 2~Myr isochrone, and 56\%--72\% of the sources are above the 1~Myr isochrone.
Including all the sources, the comparison with the PMS evolutionary tracks suggest that the cluster should be very young, i.e., younger than $\sim$1~Myr. However, we note that there are many sources located above the 0.5~Myr isochrone, which have Log~$T_{\rm eff}\leq3.49$, corresponding to stellar masses $\leq$0.15~\Msun. If excluding the sources with Log~$T_{\rm eff}\leq3.49$, a median age of $\sim$1~Myr can be derived for the cluster. 

The over-dense population with Log~$T_{\rm eff}\leq3.49$ which appear above 0.5 Myr in H-R diagram can be better fit if we use the magnetic evolutionary tracks from \cite{2016A&A...593A..99F}; see Figure~\ref{Fig:magHRD}. \cite{2016A&A...593A..99F} include the effect of magnetic fields on the stellar structure, which can inhibit the convection  and thereby slow down contraction rates and result in larger stellar radii, which then results in larger luminosities, especially for the very low-mass stars. A comparison with their isochrones suggest a median age of $\sim$2\,Myr for the Trapezium cluster.

\section{Discussion}\label{SECT:discussion}

\subsection{Luminosity Spreads}
The H-R diagrams in Fig.~\ref{Fig:HRD} show large spreads in luminosities ($\sigma$(log~$L_{\star}/L_{\odot})\sim$0.3) for sources with similar spectral types, which is much larger than that found in coeval loose associations like TW~Hya and MBM~12 \citep{2014ApJ...786...97H}. The luminosity spread in the H-R diagram has been usually attributed to an intrinsic age spread (see e.g. \citealt{2011MNRAS.418.1948J,2017ApJ...842..123F}). If we assume that the luminosity dispersion in the Trapezium cluster is only due to the age spread, there is an age dispersion of $\sim$2\,Myr in this cluster. However, other factors like veiling, binarity, starspots, and circumstellar disk orientations can contribute to the observed spread. As we have accounted for veiling in our study and still see large luminosity spreads, we can exclude that veiling contributes to most of the observed scatter.

We also argue that binaries are not the main contributor to the scatter. In an extreme case, if we assume half of the cluster population to have equal-mass binary companions, and the other half to be single stars, the luminosity dispersion would be only $\sim$0.15. Realistically, the contribution of the binaries in the luminosity spread should be much less than 0.15~dex, since the suggested companion mass ratios have a flat distribution \citep{2011ApJ...731....8K,2013MNRAS.430L...6G}. Hence, binaries are not the major cause of the luminosity spread. 

Starspots can contribute to the luminosity spread \citep{2017ApJ...836..200G}. It is well known that low-mass young stars have large spot coverage and show strong magnetic fields \citep{1996ApJ...459L..95J}. Heavily spotted models (50\% coverage) of 0.1-1.2\,\Msun\ stars could have inflated stellar radii by a factor of 10\% during the PMS stage \citep{2015ApJ...807..174S}, and induce a luminosity difference between the heavily spotted models and spot-free models by a factor of two. A precise evaluation of the contribution of the starspots in our sample requires high spectral resolution data to estimate the spot coverage as done in \cite{2017ApJ...836..200G}.

\begin{figure}
\begin{center}
\includegraphics[angle=0,width=1\columnwidth]{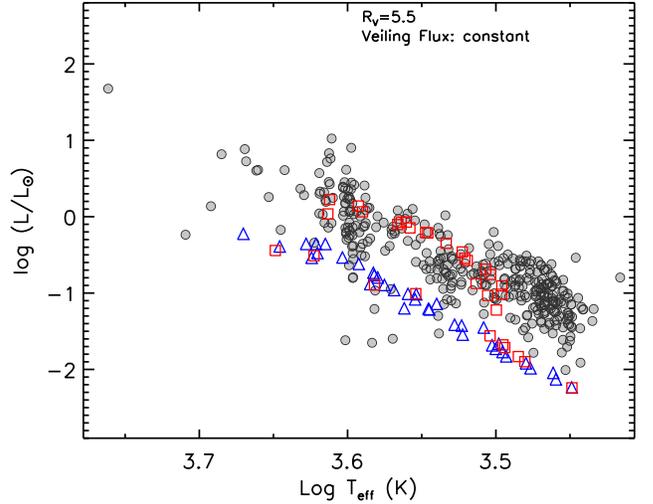}
\caption{HR diagram for the sources (grey-color filled circles) in Trapezium cluster, compared with the position of 1~Myr-old cold models (blue open triangles) and 1~Myr-old hybrid models (red open squares) from \cite{2017A&A...597A..19B}.}
\label{Fig:accHRD}
\end{center}
\end{figure}

 \begin{figure*}
\begin{center}
\includegraphics[angle=0,width=1\columnwidth]{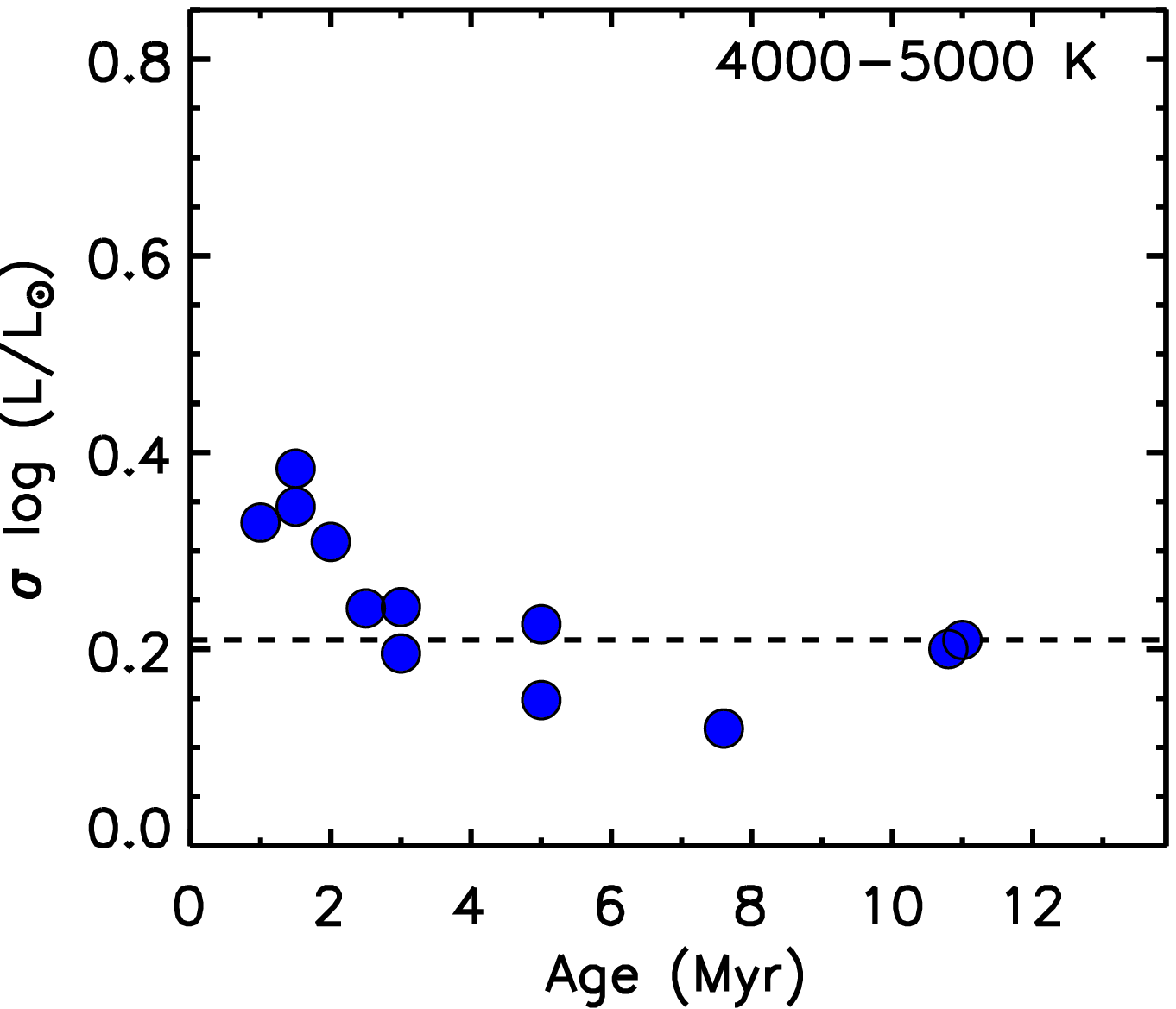}
\includegraphics[angle=0,width=1\columnwidth]{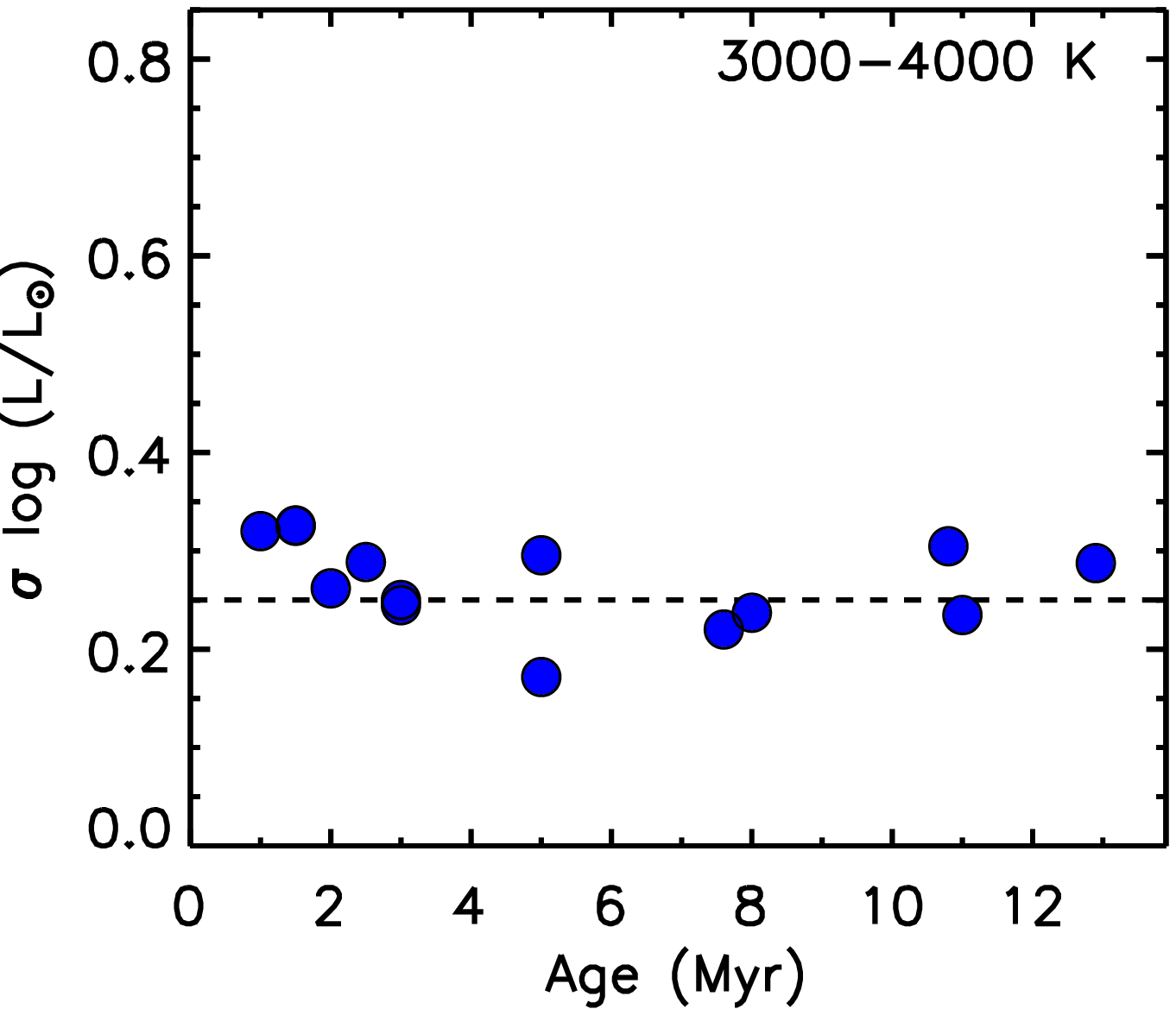}
\caption{Luminosity spreads as a function of cluster/group ages within two $T_{\rm eff}$ bins, 4000--5000~K and 3000--4000~K.}
\label{Fig:Lum_spread}
\end{center}
\end{figure*}

\setcounter{table}{1}
\begin{table*}
\renewcommand{\tabcolsep}{0.1cm}

\caption{Luminosity spreads within different temperature bins \label{Table:Lumspread}}
\begin{center}
\begin{tabular}{cccccccccccccccccc}
\hline
&  &\multicolumn{2}{c}{4000$-$5000~K} & & \multicolumn{2}{c}{3000$-$4000~K} & \\
\cline{3-4} \cline{6-7} 
Regions & Age &$\sigma($Log~$L_{\star}/L_{\odot}$) & N  & &$\sigma($Log~$L_{\star}/L_{\odot}$) & N  & References \\
\hline
Trapezium &1~Myr  &0.33  &35 & &0.32 &204 &1 \\
Taurus &1.5~Myr   &0.38  &20 & &0.33 &127 &2 \\
NGC2068/2071 &1.5~Myr  &0.34  &23 & &0.33 &105  &3 \\
Cha~I &2~Myr  &0.31  &18 & &0.26 &153  &4 \\
IC~348 &2.5~Myr  &0.24  &20 & &0.29 &203  &5 \\
NGC~2264 &3~Myr  &0.20  &181 & &0.25 &298  &6 \\
$\sigma$~Ori &3~Myr  &0.24   &20    & &0.24 &122  &7 \\
NGC~2362 &5~Myr  &0.15  &27 & &0.17 &62  &8 \\
Upper~Sco &11~Myr  &0.21  &65 & &0.23 &562  &9 \\
Ori~OB1b    &5~Myr   &0.23  &63 & &0.30 &412  &10\\
25~Ori    &7.6~Myr  &0.12     &16  & &0.22 &172  &10\\
HD~35762  &8~Myr  &    &   & &0.24 &74  &10\\
Ori~OB1a    &10.8~Myr   &0.20  &68 & &0.30 &653 &10\\
HR~1833  &12.9~Myr  &    &   & &0.29  &54   &10\\
\hline
\end{tabular}
\end{center}
References--1. This work, 2. \cite{2014ApJ...786...97H}, 3.\cite{2009A&A...504..461F}, 4. \cite{2007ApJS..173..104L}, 5. \cite{2003ApJ...593.1093L}, 6. \cite{2018A&A...609A..10V}, 7. \cite{2014ApJ...794...36H}, 8. \cite{2005AJ....130.1805D}, 9. \cite{2018AJ....156...76L}, 10. \cite{2019AJ....157...85B}
\end{table*}
\normalsize

 \begin{figure*}
\begin{center}
\includegraphics[angle=0,width=2\columnwidth]{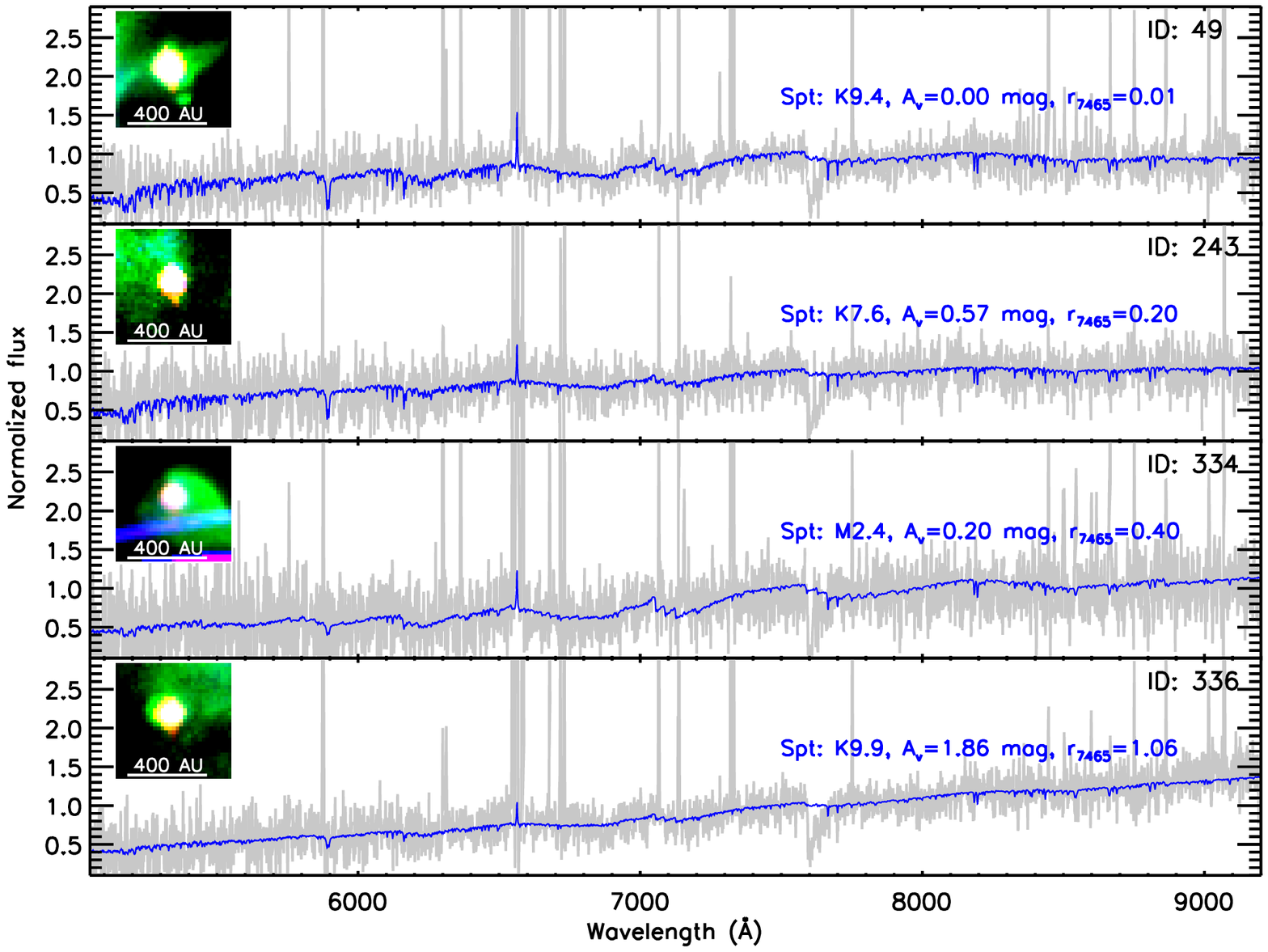}
\caption{MUSE spectra of four sub-luminous objects, numbered sources in Fig.~\ref{Fig:HRD}, are  overplotted with their best-fit spectral templates. The best-fit templates are obtained with a constant Accretion Continuum Spectrum and $R_{V}=5.5$. In each panel, the inset is the HST ACS color-composite image of the proplyd; HST image data were taken from \cite{2008AJ....136.2136R}. The tails of these four proplyds are obvious in the HST image in the F658N band.}
\label{Fig:SPT_under}
\end{center}
\end{figure*}

Theoretical studies have argued that episodic accretion during the phase of protostellar evolution can strongly affect the structure and evolution of low-mass stars during the PMS stages. It can also introduce large luminosity spreads in observed H-R diagrams without any age spreads  \citep{2010A&A...521A..44B,2012ApJ...756..118B}. \cite{2017A&A...597A..19B} performed self-consistent calculations by coupling  models of collapsing pre-stellar cores and stellar evolution models of accreting objects. In their calculation, they denote $\alpha$, the fraction of accreting internal energy absorbed by the central protostars. They consider two scenarios: (1) a cold accretion model with $\alpha$=0, meaning all accretion energy is radiated away; (2) a hybrid accretion model with $\alpha$=0, when accretion rates are smaller than a critical value ($\dot{M}_{\rm cr}$), and $\alpha$=0.2 when accretion rates are larger than a critical value.  Figure~\ref{Fig:accHRD} compares the models with our data. A striking feature is seen in the comparison: the cold accretion model fits quite well a group of sub-luminous sources, which appear below the pre-main sequence loci. In this figure, the models tend to show a bimodal distribution, since the $\dot{M}_{\rm cr}$ is fixed ($\dot{M}_{\rm cr}=10^{-5}~M$\accunit) in the calculations. A variation of 
$\dot{M}_{\rm cr}$  can fill the gap and reproduce the observed luminosity spread when Log$T_{\rm eff}\gtrsim$3.5. However, the models fail to reproduce the luminosity  spread when Log$T_{\rm eff}\lesssim$3.5.

Such luminosity spreads are commonly seen even for clusters with different ages.  Figure~\ref{Fig:Lum_spread} shows the luminosity spreads as a function of cluster/group ages within two temperature bins: 4000$-$5000~K and 3000$-$4000~K. In order to remove the contribution of the luminosity spreads from the Log$L_{\star}$--Log$T_{\rm eff}$ relation, we first fit the Log$L_{\star}$--Log$T_{\rm eff}$ relation with a linear function for each cluster within each temperature bin, and then subtract the calculated Log$L_{\star}$ at Log$T_{\rm eff}$  using the fitted linear function from observed Log$L_{\star}$. {\newrev We then calculate the standard deviation among the residual luminosity and use it as the luminosity spreads. } The plotted clusters/groups include the Trapezium cluster, the Taurus star-forming region, NGC~2068/2071, the Cha~I star forming region, IC~348, NGC~2264, $\sigma$~Ori, NGC~2362,  Upper~Sco, and Orion OB associations including Ori~OB1a, Ori~OB1b, 25~Ori, HD~35762, and HR~1833. In Table~\ref{Table:Lumspread}, we list the luminosity spread and number of stars within each temperature bin, the age of each cluster/group, as well the references. The  luminosity spreads are calculated only when the number of stars in each temperature bins are larger than 10. For the Upper~Sco and Orion OB associations, no stellar luminosities are listed in tables in \cite{2018AJ....156...76L}  and \cite{2019AJ....157...85B}. We calculate the stellar luminosity for each source using the method described in \cite{2017AJ....153..188F} taking its spectral type and extinction from the literature \citep{2018AJ....156...76L,2019AJ....157...85B}. Beside the Trapezium cluster and Taurus, the veiling effect has not been considered in deriving the spectral types and stellar luminosities. However, it should only affect the young clusters, such as NGC~2068/2071 and Cha~I where more than 50\% of sources are still accreting \citep{2010A&A...510A..72F}.

Figure~\ref{Fig:Lum_spread} shows that the large $\sigma$(log$L_{\star}/L_{\odot}$) perseveres  even in the old clusters, such as Upper~Sco within the two temperature bins.  The $\sigma$(log$L_{\star}/L_{\odot}$) tends to flatten ($\sim$0.20-0.25) after $\sim$2~Myr. If this is due to the age spread, $\sigma$(log$L_{\star}/L_{\odot}$)=0.20 at $\sim$2~Myr require an age dispersion of $\sim$2~Myr, and the same dispersion   at $\sim$11~Myr require an age dispersion of $\sim$7~Myr, according to PMS evolutionary tracks from \cite{2015A&A...577A..42B}.  {\newrev Thus, increasing age spreads with increasing cluster ages are required to reproduce the constant $\sigma$(Log$L_{\star}/L_{\odot}$).}  This argues against age spread being the main contribution of luminosity spread. {\newnewrev Figure~\ref{Fig:Lum_spread} also shows $\sim$0.1~dex larger in luminosity spreads for the younger regions, like the Trapezium cluster and Taurus, than those older regions. While the binarity induces a similar luminosity spread for clusters at different ages,  other factors, e.g., starspots, accretion history and circumstellar disk orientations, can induce a larger luminosity spread for young clusters than older ones \citep{2020ApJ...893...67M,2017A&A...597A..19B,2010A&A...521A..18G}.}


\subsection{Proplyds}\label{sect:sub_proplyd}

 Proplyds are  protoplanetary disks being photoevaporated by ultraviolet radiation from nearby massive stars and show cometary structures with tails pointing away from the massive stars \citep{1993ApJ...410..696O}. In the Orion nebula, there are more than a hundred known proplyds. In Figure~\ref{Fig:HRD} there are about ten proplyds, which appear sub-luminous compared to other young stars with the same spectral types. In particular, four proplyds (Soureces 49, 243, 334, 336) in Figure~\ref{Fig:HRD}(a,b,c) and three proplyds in Figure~\ref{Fig:HRD}(d) appear below the isochrone of 50~Myr. In Fig.~\ref{Fig:SPT_under}  we show their MUSE spectra as well their best-fit spectral templates.  We find that for each object, the best-fit template can fit the observation well. Such sub-luminous young stars are often found in young star-forming regions \citep{2009A&A...504..461F,2013ApJS..207....5F,2003A&A...406.1001C,2004ApJ...616..998W,2013A&A...549A..15F}. They are explained as young stars with highly inclined disks. The stellar light is reduced due to attenuation by the dusty disk and the observed light mainly comes from photons scattered off the disk surface.  However, we cannot exclude the possibility that  the stellar parameters of those sub-luminous objects are not well determined,  since their spectra are rather noisy.

Furthermore, we can test if there is any difference on the stellar properties between the proplyds and other sources.  It is clear that in the H-R diagram the majority of the proplyds are well-mixed among other sources; see Figure~\ref{Fig:HRD}. {\newrev The  Kolmogorov-Smirnov tests} on the distributions of the luminosity and the spectral types of the  proplyds and other young stars return a high probability (P=0.51 and 0.27 for luminosity and spectral type, respectively) for both the proplyds and other sources to be drawn from the same parent population. Since the stellar ages and masses are derived from the stellar luminosity and spectral types, we conclude that the proplyds have similar ages and show similar mass function to other sources in the cluster. This suggests that the majority of the proplyds 
formed as other stars in the cluster.

\section{summary}\label{SECT:summary}

We present a spectroscopic study of 361 young stars in the Trapezium cluster, and analyze their flux-calibrated optical spectra to derive their spectral type, extinction, and optical veiling due to the accretion simultaneously. {\newrev Based on this enhanced spectroscopic analysis, we update the H-R diagram of the Trapezium cluster, and present the main results as follows.}

\begin{itemize}
    
      \item {\newrev We improve the spectral classification of young stars in the Trapezium cluster using the X-shooter spectra of WTTSs as  templates.}
    
      \item {\rev In the Trapezium cluster, the optical total-to-selective extinction ratio ($R_{\rm V}$) is suggested to be 5.5.}

   \item {\newrev  With the derived spectral type and stellar luminosity, we present an improved H-R diagram of the Trapezium cluster. The age of the Trapezium cluster is estimated to be $\sim$1\,Myr with the non-magnetic PMS evolutionary tracks and $\sim$2\,Myr with the magnetic PMS evolutionary tracks.}
  . 
   
    \item {\newrev We find the magnetic  PMS evolutionary tracks can better explain the over-luminous low-mass young stars with Log~$T_{\rm eff}\leq3.49$ in H-R diagram than the non-magnetic PMS evolutionary tracks.} 
    \item {\newrev In the H-R diagram, there are about ten sources that appear below the 10~Myr isochrone, which can be explained by {\rev high-inclination disks} or the cold accretion model from \cite{2017A&A...597A..19B}.}  

    \item {\newrev The Trapezium cluster shows a large luminosity spread, $\sigma$(Log$L_{\star}/L_{\odot})\sim$0.3, in the H-R diagram.
        We collect a sample of 14 clusters/groups at different ages, and find that the luminosity spread remains in older ($\sim$10~Myr) cluster and the luminosity spread tend to be constant after 2~Myr. This suggests that the age spread is not the main contributor to the luminosity spread.}
    
     \item{\rev There are no significant difference on the spectral types and stellar luminosity between the prolyds and other members in the cluster.}
    
\end{itemize}

{\rev Our work further stress the importance of accounting for veiling  when deriving spectral types of accreting young stars.}

\acknowledgments
Many thanks to the anonymous referee for comments that help to improve this paper.  This material is based upon work supported by the National Aeronautics and Space Administration under Agreement No. NNX15AD94G for the program "Earths in Other Solar Systems". The results reported herein benefited from collaborations and/or information exchange within NASA Nexus for Exoplanet System Science (NExSS) research coordination network sponsored by NASA's Science Mission Directorate.

%

\vspace{5mm}
\facilities{VLT: (MUSE)}


\setcounter{table}{0}
\clearpage
\begin{longrotatetable}
\begin{deluxetable*}{rcccrcccrcccrcccrccccccccccccccccc}
\renewcommand{\tabcolsep}{0.02cm}
\tablecaption{A List of the Sources in This Work, as well as Their Spectral types and stellar luminosities\label{tab:tabe_ONC}}
\tablewidth{700pt}
\tabletypesize{\tiny}
\tablehead{
&  &  & \multicolumn{9}{c}{$R_{V}=3.1^{a}$} & \multicolumn{9}{c}{$R_{V}=5.5^{b}$}& &\\
\cline{4-12} \cline{14-22}
&   &   & \multicolumn{4}{c}{Constant} & & \multicolumn{4}{c}{BB(7000 K)} && \multicolumn{4}{c}{Constant} & & \multicolumn{4}{c}{BB(7000 K)}\\ 
\cline{4-7} \cline{9-12} \cline{14-17} \cline{19-22}
\colhead{ID}&\colhead{RA} &\colhead{DEC} &\colhead{SpT}& \colhead{Log$L_{\star}$} &\colhead{$A_{\rm V}$} &\colhead{$r_{7465}$} & &\colhead{SpT}& \colhead{Log$L_{\star}$} &\colhead{$A_{\rm V}$} &\colhead{$r_{7465}$} & &\colhead{SpT}& \colhead{Log$L_{\star}$} &\colhead{$A_{\rm V}$} &\colhead{$r_{7465}$}& &\colhead{SpT}& \colhead{Log$L_{\star}$} &\colhead{$A_{\rm V}$} &\colhead{$r_{7465}$}& \colhead{Star$^{c}$} & \colhead{SpT}\\
\colhead{}&\colhead{} & \colhead{} & \colhead{}& \colhead{($L_{\odot}$)}& \colhead{(mag)} &\colhead{}& & \colhead{}& \colhead{($L_{\odot}$)}& \colhead{(mag)} &\colhead{}& & \colhead{}& \colhead{($L_{\odot}$)}& \colhead{(mag)} &\colhead{}& &\colhead{}& \colhead{($L_{\odot}$)}& \colhead{(mag)} &\colhead{}&\colhead{}& \colhead{(Literature\tablenotemark{d})}}
\startdata
      1& 05 35 05.37&$-$05 24 10.6&K7.6&$-0.347$&3.40&0.19&&K7.1&$-0.246$&3.64&0.10&&K7.6&$-0.042$&4.03&0.12&&K7.0&$ 0.033$&4.16&0.02&    265         &M0(H97)\\
      2& 05 35 05.40&$-$05 24 15.1&M4.5&$-0.983$&2.88&0.23&&M4.3&$-0.922$&3.06&0.15&&M4.3&$-0.776$&3.34&0.21&&M4.1&$-0.719$&3.49&0.13&    266         &M1.5(H13)M2.5(H13)\\
      3& 05 35 05.43&$-$05 25 22.1&M5.9&$-1.315$&1.86&0.02&&M6.0&$-1.303$&1.89&0.02&&M5.9&$-1.195$&2.09&0.02&&M5.9&$-1.150$&2.27&0.02&    267         &K3e(H13)\\
      4& 05 35 05.45&$-$05 22 30.4&M7.4&$-1.396$&1.88&0.35&&M7.1&$-1.216$&2.61&0.29&&M7.4&$-1.242$&2.25&0.35&&M7.1&$-1.031$&2.97&0.26&   5096         &M6(H13)M6.5(H97)M7.25(Luc01)M6.75(RRL)>M9.5(WLR)\\
      5& 05 35 05.45&$-$05 25 55.8&M5.3&$-1.319$&0.39&0.09&&M5.2&$-1.272$&0.54&0.06&&M5.3&$-1.287$&0.46&0.09&&M5.2&$-1.227$&0.64&0.06&    270         &M3.5(H97)\\
      6& 05 35 05.64&$-$05 25 19.3&K1.8&$ 0.662$&0.36&0.40&&K1.5&$ 0.817$&1.00&0.45&&K1.8&$ 0.694$&0.43&0.40&&K1.6&$ 0.906$&1.18&0.43&    273         &K0(WSH)K0-K3e(H97)A0V+K1V(?)A0V(vA)K0:e(J)K4eIV(W)K3(CK)K2(Ste)\\
      7& 05 35 05.68&$-$05 23 45.4&M2.5&$-1.454$&2.59&0.34&&M2.2&$-1.299$&3.02&0.21&&M2.5&$-1.250$&3.05&0.34&&M1.9&$-1.086$&3.45&0.17&    274         &M5::(H13)<M0e(H97)\\
      8& 05 35 05.68&$-$05 25 04.2&M4.9&$-0.616$&0.00&0.00&&M4.9&$-0.615$&0.00&0.00&&M4.9&$-0.617$&0.00&0.00&&M4.9&$-0.616$&0.00&0.00&                &\\
\enddata
\end{deluxetable*}
\tablenotetext{a}{Fitting the MUSE spectra using the extinction law from \citet{1989ApJ...345..245C} with $R_{\rm V}=$3.1. Constant: During the fitting,  Accretion Continuum Spectrum is taken as a constant. BB(7000 K):  During the fitting,  Accretion Continuum Spectrum  is assumed to be a black body emission with $T=7000$~K.}
\tablenotetext{b}{Same as $^{a}$ but for $R_{\rm V}=$5.5.}
\tablenotetext{c}{Star identifier in \cite{2013AJ....146...85H}}
\tablenotetext{d}{All the spectral types are collected from \cite{2013AJ....146...85H}.  H97: \cite{1997AJ....113.1733H}; H13: \cite{2013AJ....146...85H}; Luc01: \cite{2001MNRAS.326..695L}; RRL: \cite{2007MNRAS.381.1077R}; WLR: \cite{2009MNRAS.392..817W}; WSH: \cite{2004ApJ...601..979W}; vA: \cite{1988AJ.....95.1744V}; J: \cite{1965ApJ...142..964J}; W: \cite{1983ApJ...271..642W}; CK: \cite{1979ApJS...41..743C}; Ste: H C. Stempels 2008, private communication, high dispersion spectra, taken from \cite{2013AJ....146...85H}; SHC-ir: infrared spectral types from \cite{2004ApJ...610.1045S}; Sam: A.E. Samuel 1993 unpublished PhD thesis, taken from \cite{2013AJ....146...85H}; E: \cite{1993AJ....106..372E}; P: Prosser \& Stauffer 1995, private communication, taken from \cite{2013AJ....146...85H}; HT: \cite{1986ApJ...307..609H} or reference therein; Ham:  C. Hamilton 1994 unpublished masters thesis, taken from \cite{2013AJ....146...85H}; LR: \cite{2000ApJ...540.1016L}; Sta: K. Stassun 2005, private communication, low dispersion spectra,taken from \cite{2013AJ....146...85H}; Par: \cite{1954TrSht..25....1P}; Dae: \cite{2012A&A...540A..46D}; CK: \cite{1979ApJS...41..743C}; HB: \cite{2002A&A...396..513H}; GS: \cite{1946PASP...58..366G}; SHC: optical spectral types from \cite{2004ApJ...610.1045S}}
\end{longrotatetable}


\appendix

\section{Flux re-calibration}\label{Appen:cali}
\begin{figure*}
\begin{center}
\includegraphics[width=0.48\columnwidth]{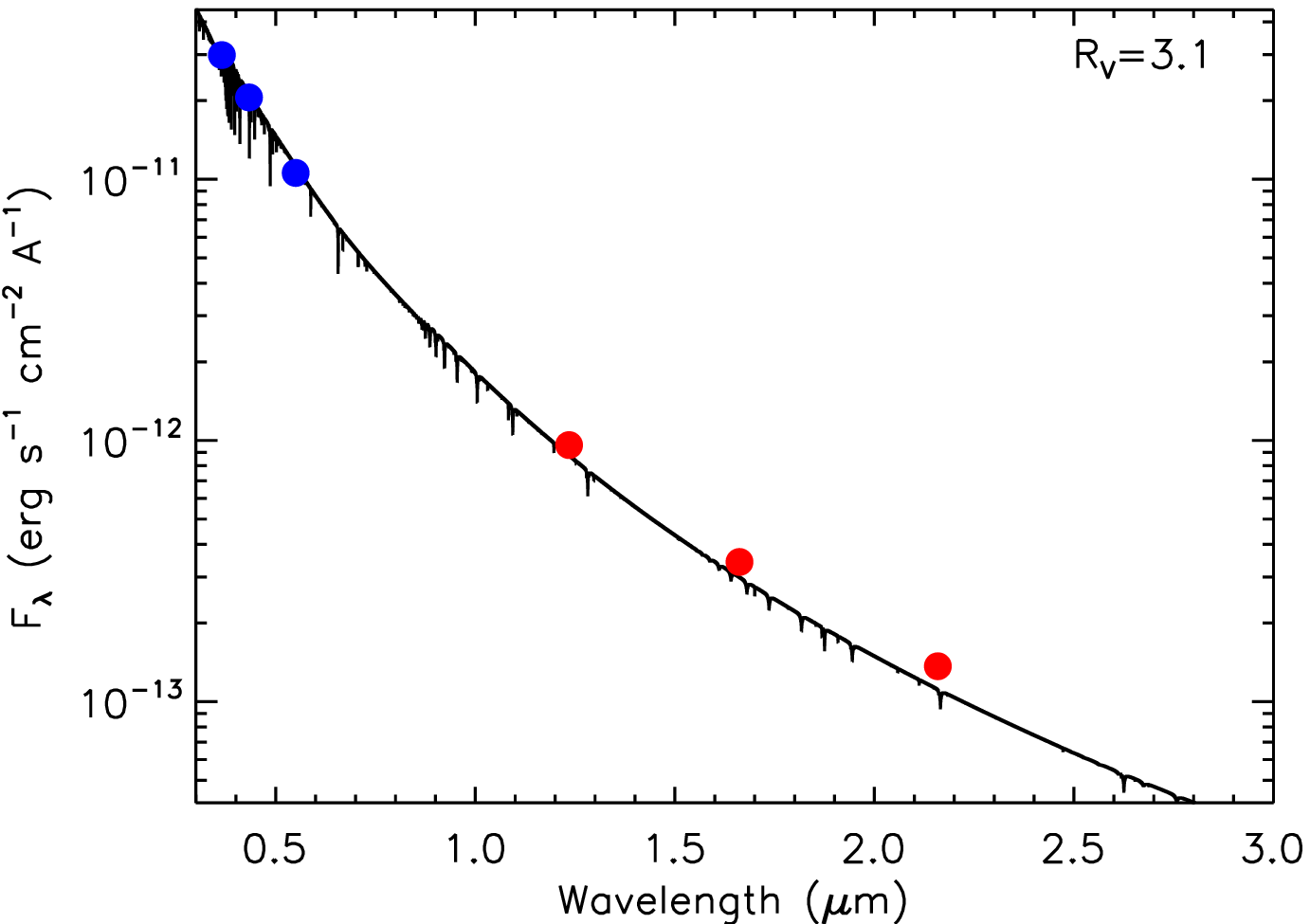}
\includegraphics[width=0.48\columnwidth]{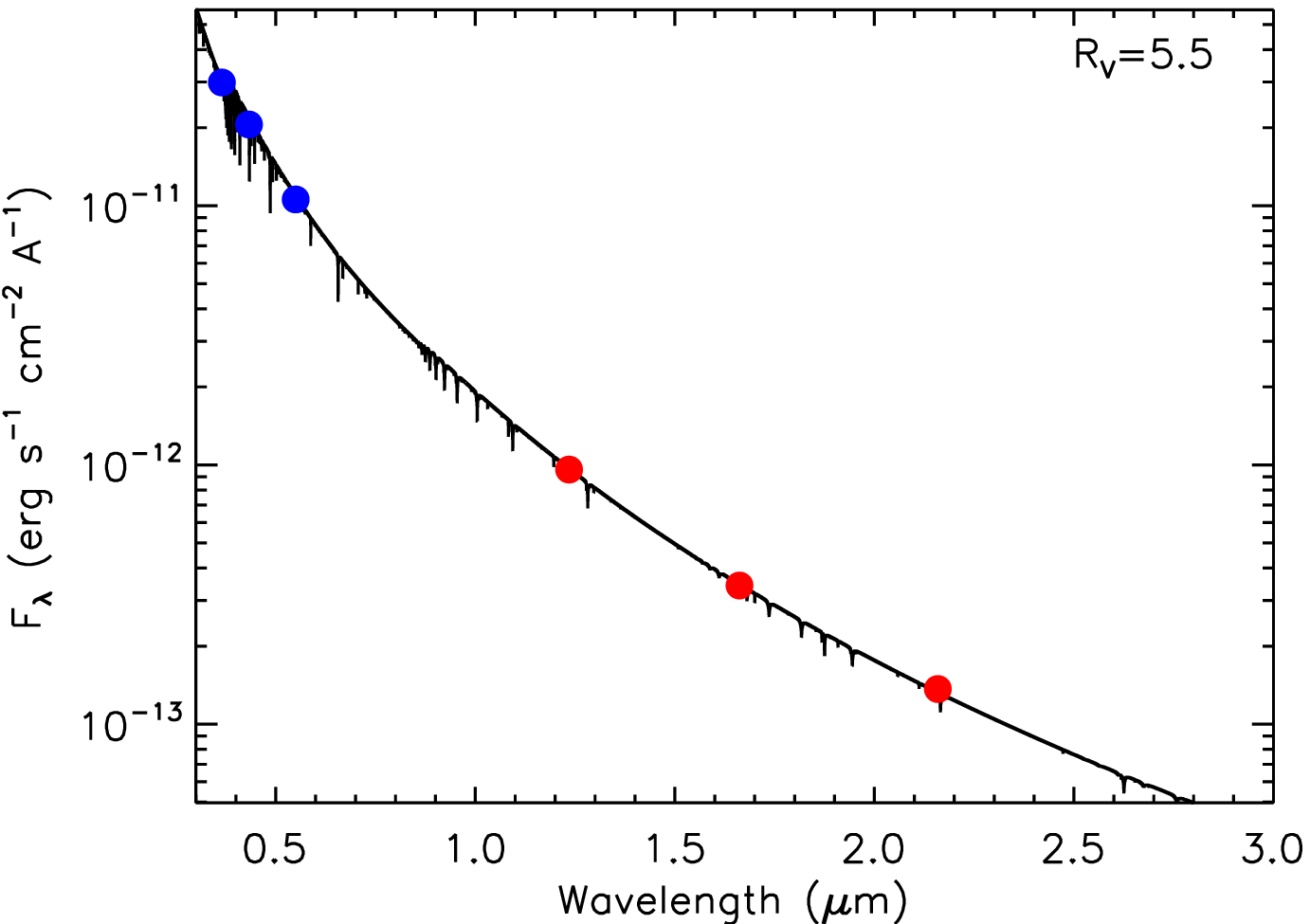}
\caption{SED of HD~37042 in  analyzed our sample in this paper. Filled circles are photometric data from $UBV$ bands (blue, \citealt{1980BICDS..19...61N}) and 2MASS $JHK_{s}$ bands (red, \citealt{2006AJ....131.1163S}). The  $UBV$ band photometry are just used for plotting, and have not been used for the SED fitting. The dark solid line show best-fit model atmosphere with $R_{V}$=3.1 (left panel) and 5.5 (right panel). }\label{Fig:SED12}
\end{center}
\end{figure*}

We refine the calibration of the MUSE data using the O-type star HD~37042 in the field.  We obtain the BOSZ Kurucz model atmosphere \citep{2012AJ....144..120M} with $T_{eff}$ from 8000~K to 35000~K, and surface gravity $log~g$ from 3.5 to 5.0, and fit the MUSE spectra of HD~37042  using the model atmospheres which are rotationally broadened and  degraded to the MUSE spectral resolution. The best-fit model is obtain by minimizing the $\chi^2$, and give $T_{\rm eff}$=29000~K and $log~g$=4.5. We then fit the broad-band photometry from {\it Gaia} and 2MASS of HD~37042  using the aforementioned best-fit model with two free parameters, extinction and stellar angular radius as in \cite{2009A&A...504..461F} and \cite{2013ApJS..207....5F}. We  use the extinction law from  extinction law of \cite{1989ApJ...345..245C}, adopting two total to selective extinction values ($R_{V}$=3.1 and 5.5). The best-fit model atmospheres with $R_{V}$=3.1 and $A_{V}$=0.76 or $R_{V}$=5.5 and $A_{V}$=1.03  are show in Fig.~\ref{Fig:SED12}. The one with $R_{V}$=5.5 can fit the broad-band photometry from optical bands to near-infrared bands, and used for the flux-calibration correction. 

\begin{figure*}
\begin{center}
\includegraphics[width=\columnwidth]{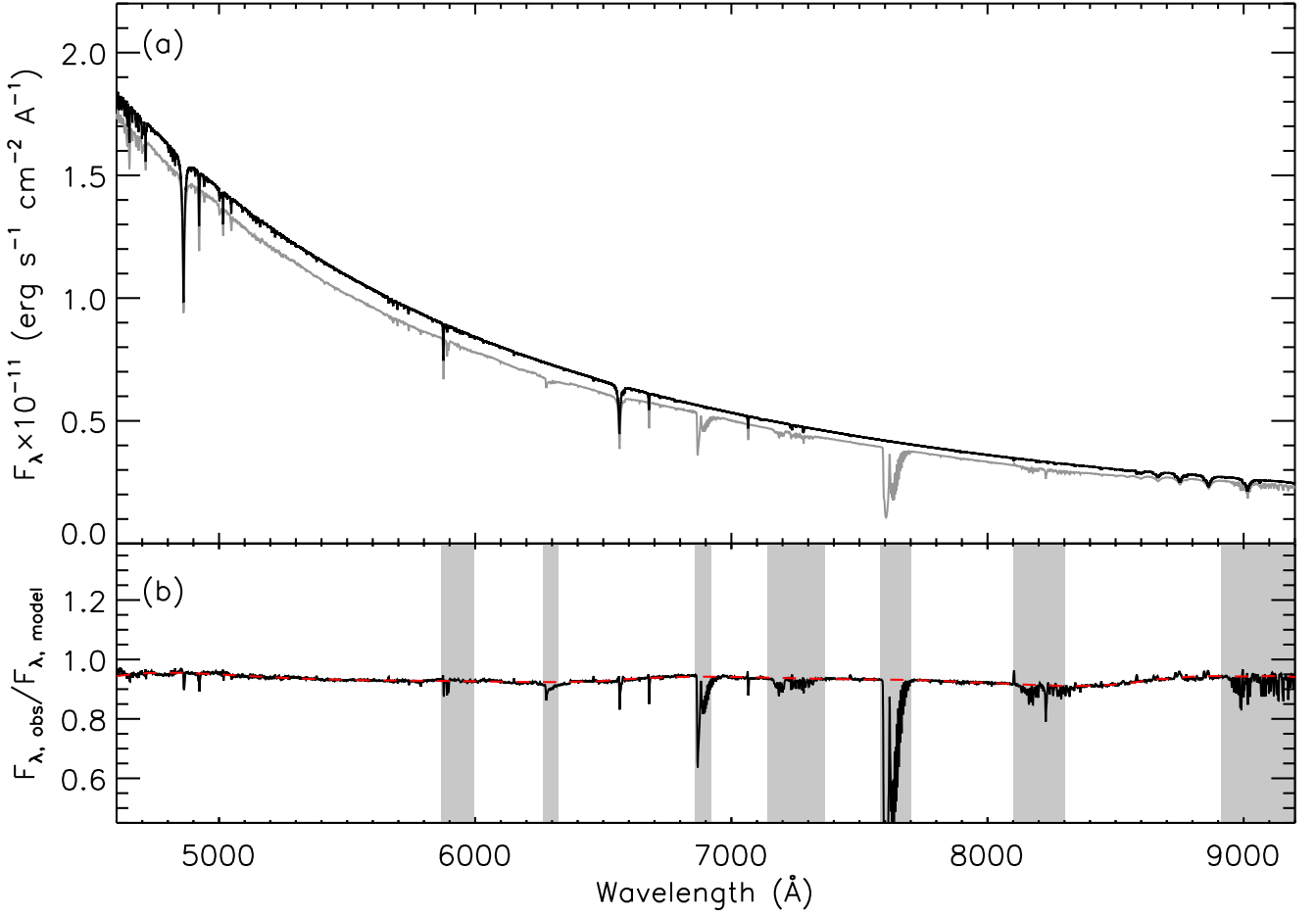}
\caption{(a) Comparison of the best-fit model (black) with the MUSE spectrum (gray) of HD~37042. (b) Flux ratio between the MUSE spectrum of HD~37042 and the best-fit model. The red dash line is used for the flux-calibration correction for the MUSE spectra of our targets. The gray color filled regions mark the wavelength ranges which are affected by the telluric absorption. }\label{Fig:fluxcali}
\end{center}
\end{figure*}

In Fig.~\ref{Fig:fluxcali}, we compare the best-fit model with $R_{V}$=5.5 with the MUSE spectra of HD~37042. The comparison show that the MUSE spectra calibrated in \citep{2015A&A...582A.114W} underestimate the flux by $\sim$5\%. We fit the flux ratio between the MUSE spectra and the best-fit model atmosphere with 5-order poly function, avoiding the regions that we affected by the telluric absorption, and use it as a correction for the MUSE spectra of our targets; see Fig.~\ref{Fig:fluxcali}(b). During the fitting process, we avoid the regions that are affected by the telluric absorption.

 \section{Pre-main sequence spectral templates}\label{Appen:template}

\begin{figure*}
\begin{center}
\includegraphics[angle=0,width=\columnwidth]{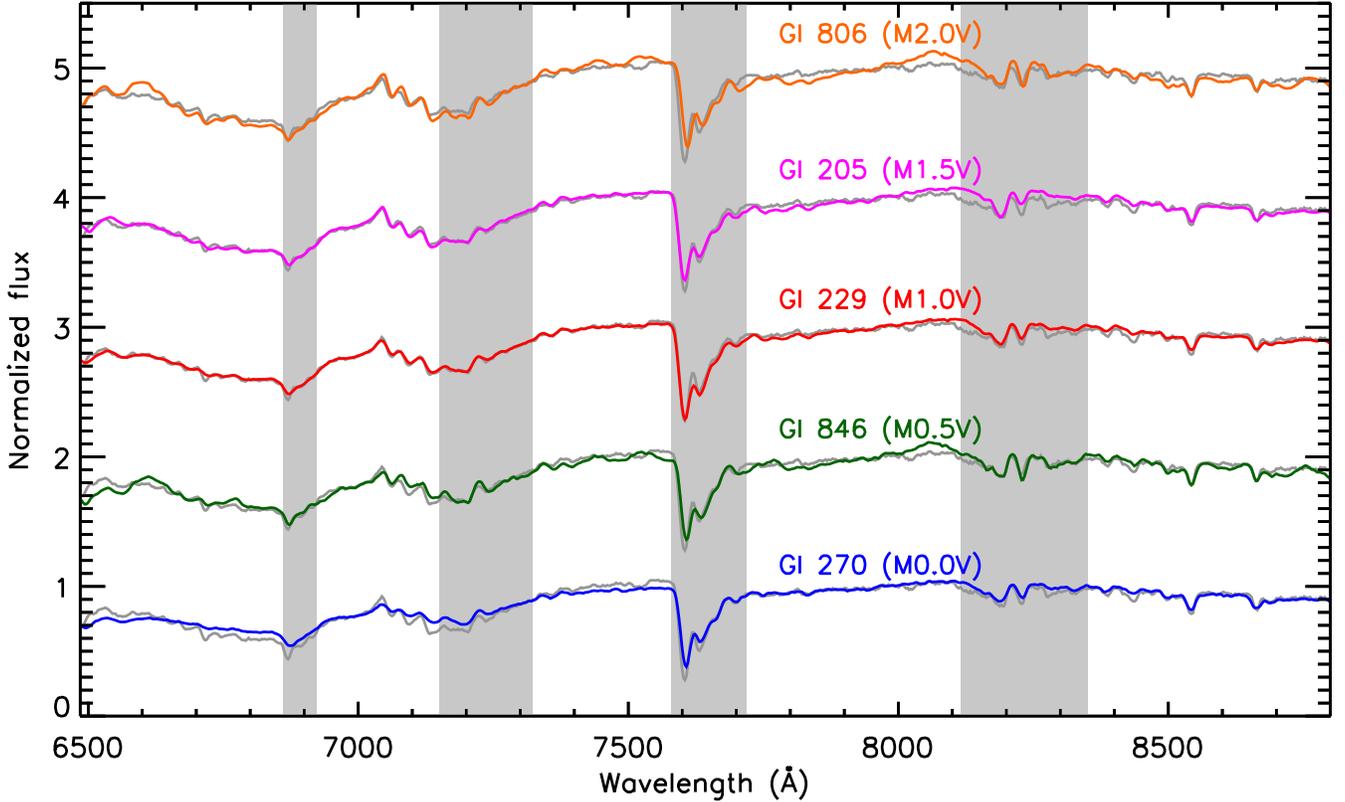}
\caption{A comparison of the spectrum of Gl~701 (gray) with those from other M-type stars. {\newnewrev In order to correct for the uncertainty in the flux calibration for each spectrum, we obtain the ratios between the spectra of Gl~701 and each comparison star at different wavelengths, and then fit the ratios using 3-order polynomial functions. The spectrum of each comparion star is multiplied by the best-fit polynomial function, and shown in the figure.}   The gray color filled regions mark the wavelength ranges which are affected by the telluric absorption. Gl~701 has been used as a M0.0V template in \cite{2013ApJS..208....9P}, but should be around M1.0V-M1.5V type.}
\label{Fig:GL701}
\end{center}
\end{figure*}

{\newrev In Table~\ref{Tab:source_Xshooter}, we list the WTTSs with X-shooter spectra and their spectral types in the literature. In this work, we mainly refer to the results from Pecaut \& Mamajek  \citep{2013ApJS..208....9P,2016MNRAS.461..794P}, Manara \citep{2013A&A...551A.107M,2017A&A...605A..86M}, and Luhman \citep{2004ApJ...609..917L, 2007ApJS..173..104L, 2017AJ....153...46L}. In the following, we discuss the techniques for the spectral classification in them.

  \cite{2013ApJS..208....9P,2016MNRAS.461..794P} classify the spectra using  spectral standards for main-sequence stars and giants and their G/K types are tied closely to the classification system of \cite{1989ApJS...71..245K} and their M types are consistent with the classification system of \cite{1991ApJS...77..417K}. Among their spectral standards, we note one star GJ~701. This star is assigned as an M0.0~V in \citep{2013ApJS..208....9P}. In Figure~\ref{Fig:GL701}, we compare the spectrum of GJ~701 with those of other M-type from \cite{1991ApJS...77..417K}. The comparisons suggest that GJ~701 should be around M1.0V-M1.5V. This explains why RECX~4 is classified as M0 in \citep{2013ApJS..208....9P} and M1.75 in \cite{2004ApJ...609..917L}. For the spectral type earlier than M0 and later than M1, the spectral types from \citep{2013ApJS..208....9P} are generally consistent with the results in the literature.
  \cite{2004ApJ...609..917L, 2007ApJS..173..104L, 2017AJ....153...46L} construct a spectral type sequence for young M dwarfs. For spectral types earlier than M5, they based on the Kirkpatrick classification system, and for the ones later than M5, they combine the spectra from field dwarfs and giant stars. 

  The majority of the X-shooter spectra are collected from \cite{2013A&A...551A.107M,2017A&A...605A..86M}.  In \cite{2013A&A...551A.107M}, they classify the spectra using the strengths of TiO, VO, CaH, etc. in the spectral region between 5800  and 9000\,\AA. In \cite{2017A&A...605A..86M}, they do the spectral classification based on the scheme developed in \cite{2014ApJ...786...97H} for ones earlier than K5, and the ones from \cite{2007MNRAS.376..580J} for K6 and later K type. For M types, \cite{2017A&A...605A..86M} use the mean values derived using the methods in \cite{2007MNRAS.381.1067R}, \cite{2007MNRAS.376..580J}, and \cite{2014ApJ...786...97H}. The three methods are all based on the strengths of molecular bands, and the typical differences in the spectral types from the three methods is about 0.5~subclasses.

While  most of literature SpTs for most of sources agree with each other within 1 subclass, there are some sources with very large differences in literature SpT ($>2$ subclasses), e.g. TWA~15A. To help determine the used spectral types of the sources, we re-classify the spectral types using two methods: one is from \cite{2017AJ....153..188F}, and the other is to fit the X-shooter spectra with the M-type spectral templates constructed by Luhman \citep{1999ApJ...525..466L,2003ApJ...593.1093L,2004ApJ...617.1216L,2006ApJ...645..676L}. The former \sout{one} is based one the relations between the strengths of molecular bands and the spectral types, and can provide the reliable spectral classification for late-K to M type stars. The latter \sout{one} is only for M-type stars \sout{and} to calibrate spectral types of the X-shooter spectra to the M-type spectral sequence defined by Luhman. The results from the two methods are also listed in  Table~\ref{Tab:source_Xshooter} in Appendix~\ref{Appen:template}. With the new results, we can determine the spectral types of the sources with large differences in literature SpT. For instance, TWA~14A have spectral types ranging from M1.5 to M3.5 in the literature, and with the new spectral types in this work we assign M3.25 to this source. 

For the K4-K6 type X-shooter templates, the spectral types in the literature agree with each other very well, and we assign the literature SpT to them.  For the early-K and G types, we take the spectral types which are generally consistent with the ones from \cite{2013A&A...551A.107M,2017A&A...605A..86M} and \cite{2013ApJS..208....9P}. For these sources, there could be 1-5 subclass differences in the spectral types of individual sources. The differences could be due to the different methods used in the spectral classification. Reconciliation of this discrepancy is beyond the scope of this work. Furthermore, the majority of our sample are later than K5 type. Thus, this uncertainties on the early-K and G types do not affect any main results in this work.}

\setcounter{table}{2}
\begin{table*}
\scriptsize
\renewcommand{\tabcolsep}{0.03cm}
\begin{center}
\caption{The parameters for WTTSs \label{Tab:source_Xshooter}}
\begin{tabular}{ccccc|c|c|c|c|c|ccccc} 
\hline \hline
ID  &Name &RA      &DEC     &Region   &SpT$^{\alpha}$       &SpT$^{\beta}$  &SpT$^{\gamma}$    & Other  &              &SpT$^\delta$  &SpT$^\epsilon$ &Used           &$A_{\rm V}$ \\
    &    &(J2000) &(J2000) &       &                &          &               &SpT        & Ref                    &             &  &SpT   & (mag)  \\  
\hline
1 & RECX\,1 &08 36 56.24  &$-$78 56 45.7  &$\eta$~Cha & K5             &           &K6        &K4, K4, K7, K7 &1, 2, 3, 34 &              & &K6        &0\\

2 & RECX\,3 &08 41 37.03  &$-$79 03 30.4&$\eta$~Cha  & M3.5          &           &M3.25     &M3, M3      &2, 3     &M3.2$\pm$0.3       &M3.25  &M3.25           &0  \\
3 & RECX\,4 &08 42 23.77  &$-$79 04 03.0&$\eta$~Cha  & M0              &            &M1.75    &K7, M1.5    &2, 3    &M1.1$\pm$0.6       &M1.75   &M1.75         &0  \\
4 & RECX\,6 &08 42 38.77 &$-$78 54 42.8&$\eta$~Cha  &                  &           &M3        &M2, M3       &2, 3  &M3.0$\pm$0.2     &M3.0  &M3.0       &0  \\
5 & RECX\,10&08 44 31.90  &$-$78 46 31.2&$\eta$~Cha   & K9              &            &M1       &K7/M0, K7, M0.5 &1, 2, 3 &  K9.0$\pm$1.0  &M0.5 &M0.5      &0   \\
6 & RECX\,12&08 47 56.77  &$-$78 54 53.2&$\eta$~Cha   &                 &            &M3.25    &M2, M2, M3  &1, 2, 3 &M3.3$\pm$0.3   &M3.0 &M3.25 &0    \\
7 &J0836    &08 36 10.73  &$-$79 08 18.4&$\eta$~Cha   &M5.5                 &            &M5.5     &M5.5, M5.5   &3, 4    &M5.7$\pm$0.2    &M5.5 &M5.5   &0  \\
8 &J0838    &08 38 51.50  &$-$79 16 13.7&$\eta$~Cha   &M5.25                  &          &M5.25    &M5, M5       &3, 4    &M5.5$\pm$0.1   & M5.5  & M5.25 &0  \\
9 &TWA\,2A&11 09 13.80  &$-$30 01 39.9 &TWA    &M1.5         &M2    &M2.25  &M0.5, M2, M1.5, M2.2 &5, 6, 7, 28  &M2.1$\pm$0.5    &M2.5   &M2.25          &0 \\
10 &TWA\,3B &11 10 27.81 &$-$37 31 53.2 &TWA&M4       &      &     &M4.5   &33        &M4.1$\pm$0.2    &M4.0 &M4.0   &0 \\
11 &TWA\,6 &10 18 28.70 &$-$31 50 02.9&TWA    &            &K7    &M0.0     &K7, M0,    M0.0    &5, 6, 28     &K7.2$\pm$0.5     &   &K7.0   &0 \\
12 &TWA\,7 &10 42 30.06 &$-$33 40 16.6&TWA    &M3          &M3    &M3.25     &M1, M2, M3.2       &5, 6, 28      &M3.2$\pm$0.2    & M3.25  & M3.25  &0 \\
13 &TWA\,8A &11 32 41.27&$-$26 51 56.0  &TWA&M3       &      &     &M3, M3, M3   &6, 32, 33     &M3.1$\pm$0.5        &M3.0     &M3.0   &0 \\
14 &TWA\,8B &11 30 11.94&$-$26 35 33.3    &TWA&M5.5       &      &     &M5, M5.2, M5.5, M5.5   &5, 28, 32, 33 &M5.5$\pm$0.2    &M5.5   &M5.25  &0  \\
15 &TWA\,9A&11 48 24.22 &$-$37 28 49.2&TWA    &K7            &K5    &     &K5, K5, K7, K6           &5, 6, 8, 28   &K6.3$\pm$0.7   &  &K7.0  &0\\
16 &TWA\,9B&11 48 23.73  &$-$37 28 48.5&TWA    &         &M3    &     &M1, M3.5, M3.4           &5, 8, 28      &M3.6$\pm$0.3   &M3.5  &M3.5 &0 \\
17 &TWA\,14 &11 13 26.22 &$-$45 23 42.7&TWA    &K8            &M0.5  &     &M0, M0.6, K8, M1.9       &7, 10, 28  &M0.5$\pm$0.6    &M1.25  &M0.5  &0 \\
18 &TWA\,15A&12 34 20.65 &$-$48 15 13.5&TWA   &               &M3.5  &     &M1.5                &10               &M3.3$\pm$0.4   &M3.25 &M3.25  &0  \\
19 &TWA\,15B&12 34 20.47  &$-$48 15 19.5&TWA   &                 &M3    &     &M2.2, M2            &7, 10         &M3.3$\pm$0.4   &M3.0--M3.25    &M3.25  &0 \\
20 &TWA\,25 &12 15 30.72  &$-$39 48 42.6&TWA   &K9       &M0    &M0.75     &M0, M1, M0.5            &11, 6, 28    &K9.8$\pm$0.9   &M0.25 &M0.5  &0 \\
21 &TWA\,26 &11 39 51.14 &$-$31 59 21.5 &TWA  &             &M9    &M8.5      &M8, M8, M9           &12, 13, 14   &M8.5$\pm$0.4  &M8.0     &M8.5   &0 \\
22 &TWA\,29 &12 45 14.16  & $-$44 29 07.7&TWA   &         &M9.5  &M9.25  &M9.5                  &13            &M9.2$\pm$0.4  &M8.5 &M9.5 &0 \\
23 &Sz~94  &16 07 49.60 &$-$39 04 28.8 &Lupus   &            &M4    &     &M4                &15          &M3.6$\pm$0.4   &M3.5 &M3.5    &0 \\
24 &Sz~107 &16 08 41.80 &$-$39 01 37.0 & Lupus   &            &M5.5  &      &M5.5              &15        &M5.5$\pm$0.1    &M5.5 &M5.5  &0  \\
25 &Sz\,121&16 10 12.20 &$-$39 21 18.1 & Lupus   &             &M4    &      &M3                &15        &M4.1$\pm$0.3   &M4.0--M4.25 &M4.0  &0  \\
26 &Sz\,122&16 10 16.42  &$-$39 08 05.1 & Lupus   &             &M2      &     &M2              &15         &M2.1$\pm$0.4   & M2.25 & M2.25 &0\\
27 &Par-Lup3-1 &16 08 16.03 &$-$39 03 04.3 & Lupus  &            &M6.5      &     &M7.5, M6.25    &16, 17    &M6.3$\pm$0.9     &M6.5--M7.0   &M6.5 &1.4$^{d}$\\
28 &Par-Lup3-2 &16 08 35.78 &$-$39 03 47.9 & Lupus  &          &M5      &     &M6                &16         &M5.1$\pm$0.2  &M5.0   &M5.0    &0 \\
29 &RXJ1526.0-4501 &15 25 59.65 &$-$45 01 15.7 &Lupus  &K0     &G9      &     &G7, G8, G5        &6, 18, 20     &   &  &K0.0 &0  \\
30 &RXJ1508.6-4423 &15 08 37.74 &$-$44 23 17.0&Lupus  &G6     &G8      &     &G8, G1.5, G8,      &6, 18, 20     &    &   &G6 &0\\
31 &RXJ1515.8-3331 &15 15 45.36 &$-$33 31 59.8&Lupus  &K0       &K0.5      &   &K0, K0, K1, K0             &6, 19, 20, 30  &  &    &K0.5 &0.2$^{g}$   \\
32 &RXJ1538.6-3916 &15 38 38.36 &$-$39 16 54.1&Lupus  &K4       &K4      &     &K4                   &20        &  & &K4.0   &0 \\
33 &RXJ1540.7-3756 &15 40 41.17 &$-$37 56 18.5&Lupus  &K6       &K6      &     &K6                   &20        &  & &K6.0  &0  \\
34 &RXJ1543.1-3920 &15 43 06.25 &$-$39 20 19.5&Lupus  &       &K6      &     &K6                   &20          &  & &K6.0   &0 \\
35 &SO~879 &05 39 05.42  &$-$02 32 30.3 &$\sigma$~Ori &          &K7      &    &K7, K6.5, K7               &21, 24, 31  &K7.0$\pm$0.8  & &K7.0 &0.2$^{g}$  \\
36 &SO~925 &05 39 11.41  &$-$02 33 32.8 &$\sigma$~Ori  &              &M5.5    &     &M5, M5.5             &22, 23      &M5.5$\pm$0.1  &M5.5 &M5.5 &0 \\
37 &SO~999 &05 39 20.25  &$-$02 38 25.8&$\sigma$~Ori  &               &M5.5    &     &M5.5, M4              &23, 24     &M5.4$\pm$0.1   &M5.0--M5.5 &M5.25 &0 \\
38 &SO~797 &05 38 54.92  &$-$02 28 58.4&$\sigma$~Ori &               &M4.5    &     &M4.5                   &23         &M4.7$\pm$0.2   &M4.75--M5.0 &M4.75 &0\\
39 &SO~641 &05 38 38.58  &$-$02 41 55.9&$\sigma$~Ori  &               &M5    &     &M5, M5.5              &23, 27       &M5.2$\pm$0.2  &MM5.0--5.5 &M5.25 &0 \\
40 &RXJ0438.6+1546  &04 38 39.07 &+15 46 13.6 & Taurus &              &K2    &K1     &G6, K2, K2     &25, 29, 30        & &  &K2.0   &0.2$^{f}$ \\
41 &J160550.5-253313 &16 05 50.64 &$-$25 33 13.6 &UpSco &G7           &K1        &     &K1, G7               &6, 26       & & &K1.0 &0.2$^{g}$   \\
42 &J160843.4-260216 &16 08 43.41 &$-$26 02 16.8 &UpSco &G7           &K0.5      &     &G9, G7               &6, 26       & &  &K0.0 &0.4$^{g}$   \\
43 &LM717   &11 08 02.34 &$-$76 40 34.3 &Cha~I &         &M6.5      &M6     &                     &       &M6.4$\pm$0.3    &M6.25 &M6.5 &0.3$^{e}$ \\
44 &J11195652-7504529 &11 19 56.52 & $-$75 04 52.9 &Cha~I &    &M7   &M7.25  &                     &      &M6.8$\pm$0.8    &M5.75  &M6.5  &0  \\
\hline
\end{tabular}
\end{center}
$^{\alpha}$ Spectral types are from \citet{2013ApJS..208....9P} and \citet{2016MNRAS.461..794P}; $^{\beta}$ Spectral types are from \citet{2013A&A...551A.107M} and \citet{2017A&A...605A..86M}; $^{\gamma}$ Spectral types are from \citet{2004ApJ...609..917L}, \citet{2017AJ....153...46L}, and \citet{2007ApJS..173..104L}; $^{\delta}$ Spectral types are classified using the scheme in  \citet{2017AJ....153..188F};  $^{\epsilon}$ Spectral types are classified using the spectral templates from Luhman.\\
$A_{V}$: $^{d}$ derived from $J-H$ color,  $^{e}$ derived from $J-H$ and $J-K_{S}$ colors,  $^{f}$ derived from the $r-J$ color;  $^{g}$ derived from the $V-K_{s}$ color. The used intrinsic colors are from \cite{2017AJ....153..188F} or \cite{2013ApJS..208....9P}.

\tablerefs{1. \citet{1997A&A...328..187C}, 2. \citet{1999ApJ...516L..77M}, 3. \citet{2004MNRAS.355..363L}, 4. \citet{2004ApJ...600.1016S}, 5. \citet{1999ApJ...512L..63W},  6. \citet{2006A&A...460..695T}, 7. \citet{2011ApJ...727....6S}, 8. \citet{2004ApJ...616..998W}, 9. \citet{1999A&A...346L..41S}, 10. \citet{2001ApJ...549L.233Z}, 11. \citet{2004ARA&A..42..685Z}, 12. \citet{2002ApJ...575..484G}, 13. \citet{2007ApJ...669L..97L}, 14. \citet{2008AJ....136.1290R}, 15. \citet{1994AJ....108.1071H}, 16. \citet{2003A&A...406.1001C}, 17. \citet{2014ApJ...785..159M}, 18. \citet{2002AJ....124.1670M}, 19. \citet{1997A&A...326..211W}, 20. \citet{1997A&AS..123..329K}, 21. \citet{2002A&A...384..937Z}, 22. \citet{2003A&A...404..171B}, 23. \citet{2012A&A...548A..56R}, 24. \citet{2014ApJ...794...36H}, 25. \citet{2017ApJ...838..150K}, 26. \citet{1998A&A...333..619P}, 27. \citet{2006A&A...445..143C}, 28. \citet{ 2014ApJ...786...97H}, 29. \cite{2012ApJ...745..119N}, 30. \cite{2010ApJ...724..835W}, 31. \cite{2008A&A...488..167S}, 32. \cite{2004ApJ...616..998W},  33. \cite{2019A&A...632A..46V}, 34 \cite{2006AJ....132..866R}.} 
\end{table*}
\normalsize

\end{document}